\documentclass[11pt]{article}

\usepackage{amsfonts,amssymb,chet,dsfont,graphicx,mathrsfs,pgfplots,tikz}
\allowdisplaybreaks

\newcommand{\1}{\mathds{1}}

\newcommand{\Op}[2]{\mathcal{O}_{#1}(\eta_{#2})}

\newcommand{\ee}[3]{(\eta_{#1}\cdot\eta_{#2})^{#3}}

\newcommand{\D}{\mathcal{D}}

\newcommand{\cOPE}[4]{{}_{#1}c_{#2#3}^{\phantom{#2#3}#4}}
\newcommand{\DOPE}[4]{{}_{#1}\D_{#2#3}^{\phantom{#2#3}#4}}
\newcommand{\tOPE}[6]{{}_{#1}t_{#2#3}^{#5#6#4}}

\newcommand{\Vev}[1]{\left\langle{#1}\right\rangle}

\title{Higher-Point Conformal Blocks\\in the Comb Channel}

\author{Jean-Fran\c{c}ois Fortin$^{\ast,}$\email{jean-francois.fortin@phy.ulaval.ca}, Wen-Jie Ma$^{\ast,}$\email{wenjie.ma.1@ulaval.ca} and Witold Skiba$^{\dagger,}$\email{witold.skiba@yale.edu}}

\affiliation{
$^\ast$D\'epartement de Physique, de G\'enie Physique et d'Optique\\Universit\'e Laval, Qu\'ebec, QC G1V 0A6, Canada\\
$^\dagger$Department of Physics, Yale University, New Haven, CT 06520, USA
}

\abstract{We compute $M$-point conformal blocks with scalar external and exchange operators in the so-called comb configuration for any $M$ in any dimension $d$.  Our computation involves repeated use of the operator product expansion to increase the number of external fields.  We check our results in several limits and compare with the expressions available in the literature when $M=5$ for any $d$, and also when $M$ is arbitrary while $d=1$.}

\date{November 2019} 

\begin{document}

\maketitle



\section{Introduction}\label{SecIntro}

Conformal blocks are essential ingredients for calculations of observables, that is correlation functions, in conformal field theories (CFTs).  A CFT is completely specified by its spectrum of primary operators and its operator product expansion (OPE) coefficients.  This set of numerical data is often referred to as the CFT data.  Two- and three-point functions are given pretty much directly in terms of the CFT data.  However, correlation functions with more than three points depend on the invariant cross-ratios and such dependence is encoded by the conformal blocks.  The higher-point functions are constructed from the CFT data and appropriate conformal blocks.

Even though conformal blocks are prescribed by the conformal symmetry, computing blocks is far from straightforward.  Various methods for obtaining the blocks have been developed over the years.  These include solving the Casimir equations \cite{Dolan:2003hv,Dolan:2011dv,Kravchuk:2017dzd}, the shadow formalism \cite{Ferrara:1972xe,Ferrara:1972uq,SimmonsDuffin:2012uy}, weight-shifting \cite{Karateev:2017jgd,Costa:2018mcg}, integrability \cite{Isachenkov:2016gim,Schomerus:2016epl,Schomerus:2017eny,Isachenkov:2017qgn,Buric:2019dfk}, 
utilizing the AdS/CFT correspondence \cite{Hijano:2015zsa,Nishida:2016vds,Castro:2017hpx,Dyer:2017zef,Chen:2017yia,Sleight:2017fpc}, and using the OPE \cite{Ferrara:1971vh,Ferrara:1971zy,Ferrara:1972cq,Ferrara:1973eg,Ferrara:1973vz,Ferrara:1974nf,Dolan:2000ut,Fortin:2016lmf,Fortin:2016dlj,Comeau:2019xco,Fortin:2019fvx,Fortin:2019dnq,Fortin:2019xyr,Fortin:2019pep,Fortin:2019gck}.  Various additional results for conformal blocks can be found in \cite{Giombi:2011rz,Costa:2011mg,Costa:2011dw,Costa:2014rya,Echeverri:2015rwa,Rejon-Barrera:2015bpa,Penedones:2015aga,Iliesiu:2015akf,Echeverri:2016dun,Costa:2016hju,Costa:2016xah,Gliozzi:2017hni,Pasterski:2017kqt,Faller:2017hyt,Rong:2017cow,Chen:2017xdz,Sleight:2018epi,Kobayashi:2018okw,Bhatta:2018gjb,Lauria:2018klo,Gromov:2018hut,Zhou:2018sfz,Kazakov:2018gcy,Li:2019dix} and \cite{Kos:2013tga,Hogervorst:2013sma,Dymarsky:2017xzb,Cuomo:2017wme,Dymarsky:2017yzx,Erramilli:2019njx}.

The vast majority of the results on conformal blocks, mentioned above, are devoted to four-point blocks as these have been used in the conformal bootstrap program \cite{Ferrara:1973yt,Polyakov:1974gs} and more recently for example in \cite{Rattazzi:2008pe,Poland:2018epd,ElShowk:2012ht}.  The four-point blocks are also the simplest.  So far, the only results for more than four points are in \cite{Alkalaev:2015fbw,Rosenhaus:2018zqn,Goncalves:2019znr,Parikh:2019ygo,Jepsen:2019svc}.  In \cite{Rosenhaus:2018zqn} $M$-point scalar blocks are derived in $d=1,2$ while for arbitrary $d$ a five-point block is computed.  These blocks are obtained in a specific configuration, termed the comb channel in \cite{Rosenhaus:2018zqn}.  We examine the same channel here.  Five-point blocks are also obtained in \cite{Goncalves:2019znr,Parikh:2019ygo}.  Higher-point correlation functions are interesting because of the AdS/CFT correspondence as blocks correspond to AdS diagrams.  Higher-point blocks may also be useful for the conformal bootstrap program where it may be possible to explore unitarity in higher-point channels and perhaps to reformulate bootstrap equations for spinning particles in terms of higher-point functions with scalars.

In this article, we derive $M$-point blocks of scalar operators in the comb channel by using the OPE in the embedding space.  We rely on the method developed in \cite{Fortin:2016lmf,Fortin:2019fvx,Fortin:2019dnq}.  The OPE relates an $M$-point block to an $(M-1)$-point one, so one can recursively build up higher-point functions starting from the ones with fewer points.  The action of the OPE on the most general expression that can appear in an $M$-point block has been explicitly computed in \cite{Fortin:2019fvx,Fortin:2019dnq}.  The OPE used there was formulated using a convenient choice of a differential operator in the embedding space that made calculations manageable.  This formalism allows for treatment of operators in non-trivial Lorentz representations, but here we consider only scalar operators in both external and exchange positions.  Examples of four-point conformal blocks containing operators with spin derived using the OPE formalism are in \cite{Fortin:2019gck}.

Since the action of the OPE on correlation functions with arbitrary $M$ is known the explicit application of the OPE is no longer needed.  We employ the prescription from \cite{Fortin:2019dnq} to obtain recursion relations that lead to conformal blocks.

This article is organized as follows.  We start in Section \ref{SecCB} by reviewing the most important features of the OPE approach.  We then discuss our choice of conformal invariants for $M$-point blocks.  There are $M (M-3)/2$ independent cross-ratios, but their choice is not unique.  We choose the cross-ratios such that when some external coordinates coincide, usually referred to as the OPE limit, relations between the cross-ratios in such limits are simple.  The recurrence relation that leads to $M$-point blocks and its solution are presented in Section \ref{SecF}.  In Section \ref{SecChecks}, we perform consistency checks on our result in several ways.  Such checks are important as both a validation of the method in \cite{Fortin:2019dnq}, that has not been used beyond four-points previously, and also to simply verify non-trivial algebra.  First, we confirm that when the dimension of an external operator is taken to zero, corresponding to substituting such an operator with the identity, $M$-point result reduces to that for $(M-1)$-points.  Next, we verify that for $M=5$ our results match those in \cite{Rosenhaus:2018zqn}.  Lastly, we consider our results in the $d=1$ limit, in which case there are only $M-3$ independent cross-ratios.  Again, we find agreement with \cite{Rosenhaus:2018zqn}.  We conclude in Section \ref{SecConc} while Appendixes \ref{SAppNotation}, \ref{SAppOPE}, \ref{SAppRR}, and \ref{SAppUnit} expand on some proofs and technicalities.


\section{Higher-Point Conformal Blocks}\label{SecCB}

Starting from any correlation function, it is technically straightforward to use repetitively the OPE to compute the necessary higher-point conformal blocks appearing in the correlation functions.  After a quick review of the OPE, this section introduces the scalar higher-point conformal blocks in the comb channel up to a function that will be determined in the next section.


\subsection{\texorpdfstring{$M$}{M}-Point Correlation Functions from the OPE}

In \cite{Fortin:2019dnq}, the general OPE in embedding space was determined to be
\eqn{
\begin{gathered}
\Op{i}{1}\Op{j}{2}=\sum_k\sum_{a=1}^{N_{ijk}}\cOPE{a}{i}{j}{k}\DOPE{a}{i}{j}{k}(\eta_1,\eta_2)\Op{k}{2},\\
\DOPE{a}{i}{j}{k}(\eta_1,\eta_2)=\frac{1}{\ee{1}{2}{p_{ijk}}}(\mathcal{T}_{12}^{\boldsymbol{N}_i}\Gamma)(\mathcal{T}_{21}^{\boldsymbol{N}_j}\Gamma)\cdot\tOPE{a}{i}{j}{k}{1}{2}\cdot\D_{12}^{(d,h_{ijk}-n_a/2,n_a)}(\mathcal{T}_{12\boldsymbol{N}_k}\Gamma)*,\\
p_{ijk}=\frac{1}{2}(\tau_i+\tau_j-\tau_k),\qquad h_{ijk}=-\frac{1}{2}(\chi_i-\chi_j+\chi_k),\\
\tau_{\mathcal{O}}=\Delta_{\mathcal{O}}-S_{\mathcal{O}},\qquad\chi_{\mathcal{O}}=\Delta_{\mathcal{O}}-\xi_{\mathcal{O}},\qquad\xi_{\mathcal{O}}=S_{\mathcal{O}}-\lfloor S_{\mathcal{O}}\rfloor,
\end{gathered}
}[EqOPE]
where $\eta_i$ are the embedding space coordinates, $\Delta_{\mathcal{O}}$ the scaling dimension, and $S_{\mathcal{O}}$ the operator spin defined as half the number of spinor indices.  The Lorentz quantum numbers of operators are encoded in the so-called half-projectors $(\mathcal{T}_{12}^{\boldsymbol{N}_i}\Gamma)$.  The notation is explained at great length in \cite{Fortin:2019dnq} and used to compute correlation functions up to four points in \cite{Fortin:2019xyr,Fortin:2019pep,Fortin:2019gck}.  This completely general expression simplifies significantly since we are dealing here with scalar operators thus there are no Lorentz indices to account for and only one scalar primary operator is kept in the OPE
\eqn{
\begin{gathered}
\Op{i}{1}\Op{j}{2}=\cOPE{}{i}{j}{k}\frac{1}{\ee{1}{2}{p_{ijk}}}\D_{12}^{(d,h_{ijk},0)}\Op{k}{2}+\ldots,\\
p_{ijk}=\frac{1}{2}(\Delta_i+\Delta_j-\Delta_k),\qquad h_{ijk}=-\frac{1}{2}(\Delta_i-\Delta_j+\Delta_k),
\end{gathered}
}[EqOPE-scalar]
where in the first line we omitted non-scalar primary operators and $\cOPE{}{i}{j}{k}$ is the OPE coefficient.  Further details of the notation and on the differential operator are relegated to Appendix \ref{SAppNotation}.

Obviously, arbitrary $M$-point correlation functions can be obtained from the OPE by using \eqref{EqOPE} repetitively.  For example, from the OPE $M$-point correlation functions can be computed from the $(M-1)$-point correlation functions as follows,
\eqna{
\Vev{\Op{i_1}{1}\cdots\Op{i_M}{M}}&=(-1)^{2\xi_{i_1}}\Vev{\Op{i_2}{2}\cdots\Op{i_M}{M}\Op{i_1}{1}},\\
&=(-1)^{2\xi_{i_1}}\sum_k\sum_a\cOPE{a}{i_M}{i_1}{k}\DOPE{a}{i_M}{i_1}{k}(\eta_M,\eta_1)\Vev{\Op{i_2}{2}\cdots\Op{i_{M-1}}{M-1}\Op{k}{1}},
}[EqCFOPERec]
where the sums over $k$ and $a$ collapse to just one term for scalar blocks, like in \eqref{EqOPE-scalar}.

Clearly, iterating \eqref{EqCFOPERec} $M-1$ times leads to $M$-point correlation functions written in terms of differential operators at the embedding space coordinate $\eta_1$ acting on one-point correlation functions at the same embedding space coordinate $\eta_1$, which corresponds to the comb channel (see Figure \ref{FigComb}) with each contribution to the correlation functions given by
\eqn{I_{M(\Delta_{k_1},\ldots,\Delta_{k_{M-3}})}^{(\Delta_{i_2},\ldots,\Delta_{i_M},\Delta_{i_1})}=\left.\Vev{\Op{i_2}{2}\cdots\Op{i_M}{M}\Op{i_1}{1}}\right|_{\substack{\text{proper exchanged quasi-primary}\\\text{operators with $\Delta_{k_1}$ to $\Delta_{k_{M-3}}$}}}.}[EqI]
In the following, we will focus on scalar operators only, with scalar exchange only, leading to the scalar $M$-point conformal blocks in the comb channel.
\begin{figure}[t]
\centering
\resizebox{15cm}{!}{%
\begin{tikzpicture}[thick]
\begin{scope}
\node at (-2,0) {$I_{M(\Delta_{k_1},\ldots,\Delta_{k_{M-3}})}^{(\Delta_{i_2},\ldots,\Delta_{i_M},\Delta_{i_1})}$};
\node at (0,0) {$=$};
\node at (1,0) {$\mathcal{O}_{i_2}$};
\draw[-] (1.5,0)--(10.5,0);
\node at (11.1,0) {$\mathcal{O}_{i_1}$};
\draw[-] (2.5,0)--(2.5,1) node[above]{$\mathcal{O}_{i_3}$};
\draw[-] (3.5,0)--(3.5,1) node[above]{$\mathcal{O}_{i_4}$};
\node at (6,0.6) {$\cdots$};
\draw[-] (8.5,0)--(8.5,1) node[above]{$\mathcal{O}_{i_{M-1}}$};
\draw[-] (9.5,0)--(9.5,1) node[above]{$\mathcal{O}_{i_M}$};
\node at (3,-0.5) {$\mathcal{O}_{k_1}$};
\node at (9,-0.5) {$\mathcal{O}_{k_{M-3}}$};
\end{scope}
\end{tikzpicture}
}
\caption{Conformal blocks in the comb channel.}
\label{FigComb}
\end{figure}
%


\subsection{Scalar \texorpdfstring{$M$}{M}-Point Correlation Functions in the Comb Channel}

To proceed, we first introduce the invariant cross-ratios as
\eqn{
\begin{gathered}
u_a^M=\frac{\ee{1+a}{2+a}{}\ee{3+a}{4+a}{}}{\ee{1+a}{3+a}{}\ee{2+a}{4+a}{}},\qquad1\leq a\leq M-3,\\
v_{ab}^M=\frac{\ee{2-a+b}{4+b}{}}{\ee{2+b}{4+b}{}}\prod_{1\leq c\leq a}\frac{\ee{3+b-c}{4+b-c}{}}{\ee{2+b-c}{4+b-c}{}},\qquad1\leq a\leq b\leq M-3,\\
\end{gathered}
}[EqCR]
with $\eta_{M+1}\equiv\eta_1$.  The choice \eqref{EqCR} is suggested by the OPE limits $\eta_M\to\eta_1$ and $\eta_2\to\eta_3$, which lead to the relations
\eqn{
\begin{gathered}
u_a^M\to u_a^{M-1},\qquad1\leq a\leq M-4,\\
u_{M-3}^M\to0,\\
v_{ab}^M\to v_{ab}^{M-1},\qquad1\leq a\leq b\leq M-4,\\
v_{1,M-3}^M\to1,\\
v_{a,M-3}^M\to v_{a-1,M-4}^{M-1},\qquad2\leq a\leq M-3,
\end{gathered}
}[EqM1]
when $\eta_M\to\eta_1$, and
\eqn{
\begin{gathered}
u_1^M\to0,\\
u_a^M\to\left.u_{a-1}^{M-1}\right|_{\eta_1\to\eta_1,\eta_b\to\eta_{b+1}},\qquad2\leq a\leq M-3,\\
v_{11}^M\to1,\\
v_{aa}^M\to\left.v_{a-1,a-1}^{M-1}\right|_{\eta_1\to\eta_1,\eta_b\to\eta_{b+1}},\qquad1<a\leq M-3,\\
v_{ab}^M\to\left.v_{a,b-1}^{M-1}\right|_{\eta_1\to\eta_1,\eta_c\to\eta_{c+1}},\qquad1\leq a<b\leq M-3,
\end{gathered}
}[Eq23]
when $\eta_2\to\eta_3$, respectively.  Indeed, two pairs of the external operators in the comb configuration, see Figure \ref{FigComb}, are special.  With our labelling of points these are the pairs $\mathcal{O}_{i_M},\mathcal{O}_{i_1}$ and $\mathcal{O}_{i_2},\mathcal{O}_{i_3}$ at the outer ends of the comb.  When the coordinates of either pair coincide an $M$-point function naturally reduces to an $(M-1)$-point function.  The special choice of the invariant cross-ratios above makes this limit particularly transparent and leads to the very simple expressions \eqref{EqM1G} and \eqref{Eq23G} below.  When $M=4$, the cross-ratios in \eqref{EqCR} reduce to the standard four-point cross-ratios $u$ and $v$ according to $u^4_1=v$ and $v^4_{11}=u$.

Then, the contribution of fully scalar exchanges in correlation functions \eqref{EqI} of all scalars is of the form
\eqna{
I_{M(\Delta_{k_1},\ldots,\Delta_{k_{M-3}})}^{(\Delta_{i_2},\ldots,\Delta_{i_M},\Delta_{i_1})}&=\left(\frac{\eta_{M-1,M}}{\eta_{1,M-1}\eta_{1M}}\right)^{\Delta_{i_1}/2}\left(\frac{\eta_{34}}{\eta_{23}\eta_{24}}\right)^{\Delta_{i_2}/2}\prod_{1\leq a\leq M-2}\left(\frac{\eta_{a+1,a+3}}{\eta_{a+1,a+2}\eta_{a+2,a+3}}\right)^{\Delta_{i_{a+2}}/2}\\
&\phantom{=}\qquad\times\left[\prod_{1\leq a\leq M-3}(u_a^M)^{\Delta_{k_a}/2}\right]G_M^{(d,\boldsymbol{h};\boldsymbol{p})}(\boldsymbol{u}^M,\textbf{v}^M),
}[EqCF]
where the conformal blocks are
\eqna{
G_M^{(d,\boldsymbol{h};\boldsymbol{p})}(\boldsymbol{u}^M,\textbf{v}^M)&=\sum_{\{m_a,m_{ab}\}\geq0}\frac{(p_3)_{m_1+\text{tr}_0\textbf{m}}(p_2+h_2)_{m_1+\text{tr}_1\textbf{m}}}{(p_3)_{m_1+\text{tr}_1\textbf{m}}}F_M^{(d,\boldsymbol{h};\boldsymbol{p})}(\boldsymbol{m})\\
&\phantom{=}\qquad\times\left[\prod_{1\leq a\leq M-3}\frac{(p_{a+2}-m_{a-1})_{m_a+\text{tr}_a\textbf{m}}(\bar{p}_{a+2}+\bar{h}_{a+2})_{m_a+m_{a+1}+\bar{m}_a+\bar{\bar{m}}_a}}{(\bar{p}_{a+2}+\bar{h}_{a+1})_{2m_a+\bar{m}_{a-1}+\bar{m}_a+\bar{\bar{m}}_a}}\right.\\
&\phantom{=}\qquad\times\left.\frac{(-h_{a+2})_{m_a}(-h_{a+2}+m_a-m_{a+1})_{\bar{m}_{a-1}}}{(\bar{p}_{a+2}+\bar{h}_{a+1}+1-d/2)_{m_a}}\frac{(u_a^M)^{m_a}}{m_a!}\right]\prod_{\substack{a,b\\b\geq a}}\frac{(1-v_{ab}^M)^{m_{ab}}}{m_{ab}!},
}[EqCB]
with
\eqn{
\begin{gathered}
\text{tr}_a\textbf{m}=\sum_bm_{b,a+b},\qquad\qquad\bar{m}_a=\sum_{b\leq a}m_{ba},\qquad\qquad\bar{\bar{m}}_a=\sum_{b>a}(\bar{m}_b-\text{tr}_b\textbf{m}).
\end{gathered}
}
Here, $\boldsymbol{m}$ is the vector of $m_a$ with $1\leq a\leq M-3$ which are the powers of the vector of $u_a^M$ denoted by $\boldsymbol{u}^M$.  Meanwhile, $\textbf{m}$ is the matrix of $m_{ab}$ with $1\leq a\leq b\leq M-3$ which are the powers of the matrix of $v_{ab}^M$ denoted by $\textbf{v}^M$, where it is understood that any $m_a$ or $m_{ab}$ outside these ranges are $0$.  The function $F_M^{(d,\boldsymbol{h};\boldsymbol{p})}(\boldsymbol{m})$ in \eqref{EqCB} is purely numerical in that it does not depend on the cross-ratios, it is determined below in Section \ref{SecF}.

The quantities $\boldsymbol{p}$ and $\boldsymbol{h}$, which are related to the OPE differential operators in embedding space \eqref{EqOPE}, are given explicitly in terms of the conformal dimensions by
\eqn{
\begin{gathered}
p_2=\Delta_{i_3},\qquad2p_3=\Delta_{i_2}+\Delta_{k_1}-\Delta_{i_3},\qquad2p_a=\Delta_{i_a}+\Delta_{k_{a-2}}-\Delta_{k_{a-3}},\\
2h_2=\Delta_{k_1}-\Delta_{i_2}-\Delta_{i_3},\qquad2h_a=\Delta_{k_{a-1}}-\Delta_{k_{a-2}}-\Delta_{i_{a+1}},
\end{gathered}
}
with $k_{M-2}=i_1$, $\bar{p}_a=\sum_{b=2}^ap_b$ and $\bar{h}_a=\sum_{b=2}^ah_b$.  Meanwhile, the conformal dimensions of the exchanged scalar operators are denoted by $\Delta_{k_a}$.  The general form of the conformal blocks \eqref{EqCB} is determined from the OPE limits as we argue below.


\subsection{OPE Limits}

The rationale behind the pre-factor in \eqref{EqCF}, apart from the necessary homogeneity condition, comes from the property that under the OPE limits $\eta_M\to\eta_1$ and $\eta_2\to\eta_3$, the conformal blocks transform as
\eqn{
\begin{gathered}
G_M^{(d,\boldsymbol{h};\boldsymbol{p})}(\boldsymbol{u}^M,\textbf{v}^M)\to G_{M-1}^{(d,\boldsymbol{h}';\boldsymbol{p}')}(\boldsymbol{u}^{M-1},\textbf{v}^{M-1}),\\
p'_a=p_a,\qquad2\leq a\leq M-1,\\
h'_a=h_a,\qquad2\leq a\leq M-2,
\end{gathered}
}[EqM1G]
when $\eta_M\to\eta_1$ from \eqref{EqM1} and
\eqn{
\begin{gathered}
G_M^{(d,\boldsymbol{h};\boldsymbol{p})}(\boldsymbol{u}^M,\textbf{v}^M)\to G_{M-1}^{(d,\boldsymbol{h}';\boldsymbol{p}')}\left(\left.\boldsymbol{u}^{M-1}\right|_{\eta_1\to\eta_1,\eta_a\to\eta_{a+1}},\left.\textbf{v}^{M-1}\right|_{\eta_1\to\eta_1,\eta_a\to\eta_{a+1}}\right),\\
p'_2=p_4-h_3,\qquad p'_3=\bar{p}_3+\bar{h}_3,\\
p'_a=p_{a+1},\qquad4\leq a\leq M-1,\\
h'_a=h_{a+1},\qquad2\leq a\leq M-2,
\end{gathered}
}[Eq23G]
when $\eta_2\to\eta_3$ from \eqref{Eq23}, respectively.

It is important to note that the OPE limits above, together with the limit of unit operator discussed below in Section \ref{SSecUnit}, allow the construction of the scalar $M$-point conformal blocks up to yet unspecified functions $F_M^{(d,\boldsymbol{h};\boldsymbol{p})}(\boldsymbol{m})$.  Indeed, by implementing the OPE limits and demanding that the scalar $M$-point conformal blocks match the scalar $(M-1)$-point conformal blocks with the proper parameters, the overall form of the blocks with the specific forms for the Pochhammer symbols is determined up to the function $F_M^{(d,\boldsymbol{h};\boldsymbol{p})}(\boldsymbol{m})$.  This is straightforward to verify by starting from the special $u$ cross-ratio that vanishes and the special $v$ cross-ratio that becomes one.  Indeed, these two cross-ratios must have vanishing exponents which lead to the form \eqref{EqCB} by consistency.

In summary, demanding that the conformal blocks $G$ and the function $F$ behave properly under the OPE limits and the limits of unit operator, \textit{i.e.}\ they reduce to their form with one less point, the form of the conformal blocks $G$ is settled up to the function $F$.


\section{Function \texorpdfstring{$F_M^{(d,\boldsymbol{h};\boldsymbol{p})}(\boldsymbol{m})$}{F}}\label{SecF}

Using the form \eqref{EqCB} and the OPE \eqref{EqOPE}, it is straightforward to obtain a recurrence relation for the function $F_M^{(d,\boldsymbol{h};\boldsymbol{p})}(\boldsymbol{m})$.  In this section, this recurrence relation is found, proving that the scalar $M$-point conformal blocks \eqref{EqCB} in the comb channel are correct with the function $F_M^{(d,\boldsymbol{h};\boldsymbol{p})}(\boldsymbol{m})$ given by
\eqn{F_M^{(d,\boldsymbol{h};\boldsymbol{p})}(\boldsymbol{m})=\prod_{1\leq a\leq M-4}{}_3F_2\left[\begin{array}{c}-m_a,-m_{a+1},-\bar{p}_{a+2}-\bar{h}_{a+1}+d/2-m_a\\p_{a+3}-m_a,h_{a+2}+1-m_a\end{array};1\right].}[EqF]
Note that there are $M-4$ sums in \eqref{EqF} and $F_4^{(d,\boldsymbol{h};\boldsymbol{p})}(\boldsymbol{m})=1$.  The boundary condition, which is directly obtained from the known four-point conformal blocks, will lead to the result \eqref{EqF}.


\subsection{OPE Differential Operator}\label{SecDiffOp}

From the OPE, the scalar contributions to the correlation functions are related by
\eqn{I_{M(\Delta_{k_1},\ldots,\Delta_{k_{M-3}})}^{(\Delta_{i_2},\ldots,\Delta_{i_M},\Delta_{i_1})}=\frac{1}{\eta_{1M}^{p_M}}\left(\frac{\eta_{2M}\eta_{3M}}{\eta_{1M}\eta_{23}}\right)^{h_{M-1}}\bar{\D}_{M1;23;2}^{2h_{M-1}}I_{M-1(\Delta_{k_1},\ldots,\Delta_{k_{M-4}})}^{(\Delta_{i_2},\ldots,\Delta_{i_{M-1}},\Delta_{k_{M-3}})},}
or\footnote{Note that the conformal blocks defined here are equivalent to the ones defined in \cite{Fortin:2019dnq} up to powers of $v$'s.}
\eqna{
G_M^{(d,\boldsymbol{h};\boldsymbol{p})}(\boldsymbol{u}^M,\textbf{v}^M)&=\left(\frac{\prod_{a=1}^{M-3}u_a^M}{v_{M-5,M-4}^Mv_{M-4,M-4}^M}\right)^{-\bar{p}_{M-1}-\bar{h}_{M-1}}\\
&\phantom{=}\qquad\times\bar{\D}_{M1;23;2}^{2h_{M-1}}\frac{(x_2^M)^{\bar{p}_{M-1}+\bar{h}_{M-2}}}{(1-y_{M-2}^M)^{\bar{p}_{M-2}+\bar{h}_{M-2}}(1-y_{M-1}^M)^{p_{M-1}}}G_{M-1}^{(d,\boldsymbol{h};\boldsymbol{p})}(\boldsymbol{u}^{M-1},\textbf{v}^{M-1}),
}[EqGG]
where the differential operator $\bar{\D}$ and the meaning of the subscripts $\bar{\D}_{M1;23;2}$ are reviewed in Appendix \ref{SAppNotation}.

The action of the OPE differential operator is\footnote{Note that the OPE conformal differential has been re-scaled compared to the one defined in \cite{Fortin:2019dnq}.}
\eqna{
&\bar{\D}_{M1;23;2}^{2h_{M-1}}(x_2^M)^{\bar{q}}\prod_{3\leq a\leq M-1}(1-y_a^M)^{-q_a}\\
&\qquad=(x_2^M)^{\bar{q}+h_{M-1}}\sum_{\{n_a,n_{2a},n_{ab}\}\geq0}\frac{(-h_{M-1})_{\bar{n}_2+\bar{\bar{n}}}(q_2)_{\bar{n}_2}(\bar{q}+h_{M-1})_{\bar{n}-\bar{\bar{n}}}}{(\bar{q})_{\bar{n}+\bar{n}_2}(\bar{q}+1-d/2)_{\bar{n}_2+\bar{\bar{n}}}}\\
&\qquad\phantom{=}\qquad\times\prod_{3\leq a\leq M-1}\frac{(q_a)_{n_a}}{n_{2a}!(n_a-n_{2a}-\bar{n}_a)!}(y_a^M)^{n_a}\left(\frac{x_2^Mz_{2a}^M}{y_a^M}\right)^{n_{2a}}\prod_{3\leq a<b\leq M-1}\frac{1}{n_{ab}!}\left(\frac{x_2^Mz_{ab}^M}{y_a^My_b^M}\right)^{n_{ab}},
}[EqD]
on the cross-ratios
\eqn{
\begin{gathered}
x_2^M=\frac{\eta_{1M}\eta_{23}}{\eta_{12}\eta_{3M}},\\
y_a^M=1-\frac{\eta_{1a}\eta_{2M}}{\eta_{12}\eta_{aM}},\qquad3\leq a\leq M-1,\\
z_{ab}^M=\frac{\eta_{ab}\eta_{2M}\eta_{3M}}{\eta_{23}\eta_{aM}\eta_{bM}},\qquad2\leq a<b\leq M-1.
\end{gathered}
}[EqCROPE]
In the equation \eqref{EqD} above, $\bar{q}=\sum_{a=2}^{M-1}q_a$ and
\eqn{
\begin{gathered}
\bar{n}=\sum_{3\leq a\leq M-1}n_a,\qquad\qquad\bar{n}_2=\sum_{3\leq a\leq M-1}n_{2a},\\
\bar{n}_a=\sum_{\substack{3\leq b\leq M-1\\b\neq a}}n_{ab},\qquad\qquad\bar{\bar{n}}=\sum_{3\leq a<b\leq M-1}n_{ab}.
\end{gathered}
}
The action of the OPE differential operator \eqref{EqD} on the cross-ratios \eqref{EqCROPE} was obtained in \cite{Fortin:2019dnq}.

Clearly, it is necessary to determine the cross-ratios \eqref{EqCR} in terms of the cross-ratios \eqref{EqCROPE} and vice-versa.  From their definitions, these relations are given by
\eqn{
\begin{gathered}
u_{M-4}^{M-1}=\frac{1-y_{M-1}^M}{1-y_{M-2}^M}\frac{z_{M-3,M-2}^M}{z_{M-3,M-1}^M},\\
v_{a,M-4}^{M-1}=\frac{1-y_{M-2-a}^M}{1-y_{M-2}^M}\prod_{1\leq b\leq a}\frac{z_{M-2-a+b,M-1-a+b}^M}{z_{M-3-a+b,M-1-a+b}^M},
\end{gathered}
}[Equtox]
and
\eqn{
\begin{gathered}
x_2^M=\frac{1}{v_{M-5,M-4}^Mv_{M-3,M-3}^M}\prod_{1\leq a\leq M-3}u_a,\\
y_a^M=1-\frac{v_{M-1-a,M-3}^Mv_{M-4,M-4}^M}{v_{M-2-a,M-4}^Mv_{M-3,M-3}^M},\qquad3\leq a\leq M-2,\\
y_{M-1}^M=1-\frac{v_{M-4,M-4}^M}{v_{M-3,M-3}^M},\\
z_{2a}^M=\frac{v_{a-4,a-4}^Mv_{M-5,M-4}^M}{v_{M-2-a,M-4}^M}\prod_{1\leq b\leq a-3}\frac{1}{u_b},\qquad4\leq a\leq M-1,\\
z_{3a}^M=\frac{v_{a-5,a-4}^Mv_{M-4,M-4}^M}{v_{M-2-a,M-4}^M}\prod_{1\leq b\leq a-3}\frac{1}{u_b},\qquad4\leq a\leq M-1,\\
z_{ab}^M=\frac{v_{b-a-2,b-4}^Mv_{M-5,M-4}^Mv_{M-4,M-4}^M}{v_{M-2-a,M-4}^Mv_{M-2-b,M-4}^M}\prod_{1\leq c\leq b-3}\frac{1}{u_c},\qquad4\leq a<b\leq M-1,
\end{gathered}
}[Eqxtou]
respectively.  It is now possible to obtain the recurrence relation for $F_M^{(d,\boldsymbol{h};\boldsymbol{p})}(\boldsymbol{m})$.


\subsection{Recurrence Relation}

There are several steps for finding the recurrence relation for $F_M^{(d,\boldsymbol{h};\boldsymbol{p})}(\boldsymbol{m})$ from \eqref{EqGG}.  First, the original cross-ratios \eqref{EqCR} are expressed in terms of the cross-ratios \eqref{EqCROPE} using \eqref{Equtox}.  Second, we act with the OPE differential operator as in \eqref{EqD}, but re-express the cross-ratios \eqref{EqCROPE} in terms of the original cross-ratios \eqref{EqCR} using \eqref{Eqxtou}.  Finally, we re-sum as many sums as possible\footnote{Note that all sums are of the Gauss' hypergeometric type ${}_2F_1(-m,b;c;1)=\frac{(c-b)_m}{(c)_m}$ for $m$ a non-negative integer.  This relation is always satisfied since the re-summations are always finite.} which leads to the following recurrence relation for the function $F_M^{(d,\boldsymbol{h};\boldsymbol{p})}(\boldsymbol{m})$,
\eqna{
F_M^{(d,\boldsymbol{h};\boldsymbol{p})}(\boldsymbol{m})&=\sum_{\{t_{a,M-4}\}\geq0}(-m_{M-3})_{t_{M-4,M-4}}\left[\prod_{1\leq a\leq M-4}\frac{(-t_{a,M-4})_{t_{a-1,M-4}}}{t_{a,M-4}!}\right]\\
&\phantom{=}\qquad\times\prod_{1\leq a\leq M-4}\frac{(-m_a)_{t_{a,M-4}}(-\bar{p}_{a+2}-\bar{h}_{a+1}+d/2-m_a)_{t_{a,M-4}}}{(p_{a+3}-m_a)_{t_{a,M-4}}(h_{a+2}+1-m_a)_{t_{a,M-4}}}F_{M-1}^{(d,\boldsymbol{h};\boldsymbol{p})}(\boldsymbol{m}-\boldsymbol{t}_{M-4}),
}[EqRR]
where $\boldsymbol{t}_{M-4}$ is the vector of $t_{a,M-4}$ with $1\leq a\leq M-4$.  The computation is straightforward yet long, tedious, and not really illuminating.  Hence it is only sketched in Appendix \ref{SAppOPE}.

Since the initial condition is $F_4^{(d,\boldsymbol{h};\boldsymbol{p})}(\boldsymbol{m})=1$, as obtained from the known four-point conformal blocks, \eqref{EqRR} leads to
\eqna{
F_M^{(d,\boldsymbol{h};\boldsymbol{p})}(\boldsymbol{m})&=\sum_{\{t_{ab}\}\geq0}(-m_{M-3})_{t_{M-4,M-4}}\left[\prod_{1\leq a\leq b\leq M-4}\frac{(-t_{ab})_{t_{a-1,b}}}{t_{ab}!}\right]\\
&\phantom{=}\qquad\times\prod_{1\leq a\leq M-4}\frac{(-m_a)_{t_{a-1,a-1}+\bar{t}_a}(-\bar{p}_{a+2}-\bar{h}_{a+1}+d/2-m_a)_{\bar{t}_a}}{(p_{a+3}-m_a)_{\bar{t}_a}(h_{a+2}+1-m_a)_{\bar{t}_a}}.
}[EqFp]
Here $\bar{t}_a=\sum_{b\geq a}t_{ab}$ and the summation variables $t_{ab}$ exist for $1\leq a\leq b\leq M-4$, with the ones outside this range being set to $0$.  Note that there are $(M-4)(M-3)/2$ sums in \eqref{EqFp} and $F_4^{(d,\boldsymbol{h};\boldsymbol{p})}(\boldsymbol{m})=1$ as expected.

To get to the result \eqref{EqF}, we keep the $\bar{t}_a$, which will become the summation index for the ${}_3F_2$'s appearing in \eqref{EqF}, and re-sum all the remaining $t$'s, again using Gauss' hypergeometric identities.  It is also possible to check the result directly from the recurrence relation \eqref{EqRR} as shown in Appendix \ref{SAppRR}.  This thus proves that \eqref{EqCB} with \eqref{EqF} is correct for the scalar $M$-point conformal blocks in the comb channel.


\section{Sanity Checks}\label{SecChecks}

Although the, somewhat lengthy, re-summations leading to the recurrence relation \eqref{EqRR} are straightforward, they nevertheless lead to a direct proof that the $M$-point conformal blocks in the comb channel can be written as in \eqref{EqCB}.  However, it is important to check the result in certain limits.  In this section several checks are described showing that our results are consistent.


\subsection{Limit of Unit Operator}\label{SSecUnit}

It is straightforward to check that under the limit $\Delta_{i_1}\to0$, \textit{i.e.}\ when $\Op{i_1}{1}\to\mathds{1}$ with $\Delta_{k_{M-3}}=\Delta_{i_M}$, and under the limit $\Delta_{i_2}\to0$, \textit{i.e.}\ when $\Op{i_2}{2}\to\mathds{1}$ with $\Delta_{k_1}=\Delta_{i_3}$, the $M$-point conformal blocks \eqref{EqCB} reduce to the proper $(M-1)$-point conformal blocks.  Indeed, one has
\eqn{
\begin{gathered}
G_M^{(d,\boldsymbol{h};\boldsymbol{p})}(\boldsymbol{u}^M,\textbf{v}^M)\to G_{M-1}^{(d,\boldsymbol{h}';\boldsymbol{p}')}\left(\left.\boldsymbol{u}^{M-1}\right|_{\eta_1\to\eta_M},\left.\textbf{v}^{M-1}\right|_{\eta_1\to\eta_M}\right),\\
p'_a=p_a,\qquad2\leq a\leq M-2,\\
2p'_{M-1}=\Delta_{i_{M-1}}+\Delta_{i_M}-\Delta_{k_{M-4}},\\
h'_a=h_a,\qquad2\leq a\leq M-3,\\
2h'_{M-2}=\Delta_{i_M}-\Delta_{k_{M-4}}-\Delta_{i_{M-1}},
\end{gathered}
}
when $\Op{i_1}{1}\to\mathds{1}$ and
\eqn{
\begin{gathered}
G_M^{(d,\boldsymbol{h};\boldsymbol{p})}(\boldsymbol{u}^M,\textbf{v}^M)\to G_{M-1}^{(d,\boldsymbol{h}';\boldsymbol{p}')}\left(\left.\boldsymbol{u}^{M-1}\right|_{\eta_1\to\eta_1,\eta_a\to\eta_{a+1}},\left.\textbf{v}^{M-1}\right|_{\eta_1\to\eta_1,\eta_a\to\eta_{a+1}}\right),\\
p'_2=\Delta_{i_4},\qquad2p'_3=\Delta_{i_3}+\Delta_{k_2}-\Delta_{i_4},\\
p'_a=p_{a+1},\qquad4\leq a\leq M-1,\\
2h'_2=\Delta_{k_2}-\Delta_{i_4}-\Delta_{i_3},\\
h'_a=h_{a+1},\qquad3\leq a\leq M-2,
\end{gathered}
}
when $\Op{i_2}{2}\to\mathds{1}$, respectively.  Thus, the correlation function $I_M$ reduces directly to the proper contributions to the correlation function $I_{M-1}$.  These relations are expected to be correct directly from the OPE and were indeed used to fix the form of the Pochhammer symbols in \eqref{EqCB}, up to the function $F_M^{(d,\boldsymbol{h};\boldsymbol{p})}(\boldsymbol{m})$.  For example, by setting the proper conformal dimension to zero and relating the other conformal dimensions accordingly, one of the parameters in $\boldsymbol{p}$ or $\boldsymbol{h}$ vanishes and some of the Pochhammer symbols in the original $M$-point conformal blocks constrain the powers of the appropriate cross-ratios to vanish, leading to the $(M-1)$-point conformal blocks, as required.

Although it is not as simple, it is also possible to verify that the limit $\Delta_{i_{M-1}}\to0$ which corresponds to $\Op{i_{M-1}}{M-1}\to\mathds{1}$ reduces to the proper $(M-1)$-point conformal blocks.  As before, such a proof necessitates standard Gauss' hypergeometric type re-summations, but this time it intertwines the sums from $F_M^{(d,\boldsymbol{h};\boldsymbol{p})}(\boldsymbol{m})$ with the sums from $G_M^{(d,\boldsymbol{h};\boldsymbol{p})}(\boldsymbol{u}^M,\textbf{v}^M)$.  Again, this proof is straightforward yet long and tedious, and as such it is not shown in its entirety here (it is sketched in Appendix \ref{SAppUnit}).  Although we did not verify it explicitly, it is reasonable to argue that all the extra $M-4$ sums in the function $F_M^{(d,\boldsymbol{h};\boldsymbol{p})}(\boldsymbol{m})$ are required, and that they allow the proper reduction of the $M$-point conformal blocks to the $(M-1)$-point conformal blocks when any of the external operator is set to the unit operator.  Indeed, from the limit of unit operator, there are four operators for which the limit is trivial, originating from the two associated OPEs.\footnote{The four operators are $\Op{i_2}{2}$ and $\Op{i_3}{3}$ from the left endpoint of the comb and $\Op{i_M}{M}$ and $\Op{i_1}{1}$ from the right endpoint of the comb, as shown in Figure \ref{FigComb}.}  Hence, one is left with $M-4$ operators where the limit of unit operator is not straightforward.  From the computation above (sketched in Appendix \ref{SAppUnit}) where the sums from $F_M^{(d,\boldsymbol{h};\boldsymbol{p})}(\boldsymbol{m})$ and $G_M^{(d,\boldsymbol{h};\boldsymbol{p})}(\boldsymbol{u}^M,\textbf{v}^M)$ intertwine when $\Op{i_{M-1}}{M-1}\to\mathds{1}$, it seems logical to argue that the remaining limits of unit operator $\Op{i_j}{j}\to\mathds{1}$ for $4\leq j\leq M-1$ necessitate one extra summation each.  Moreover, since there is no real difference in $G_M^{(d,\boldsymbol{h};\boldsymbol{p})}(\boldsymbol{u}^M,\textbf{v}^M)$ when $\Op{i_j}{j}\to\mathds{1}$ for any $j$ such that $4\leq j\leq M-1$, the proof should be analog to the one for $\Op{i_{M-1}}{M-1}\to\mathds{1}$.  Hence, there would be a minimum of $M-4$ extra sums necessary for $M$-point conformal blocks, irrespective of the choice of cross-ratios, as found in $F_M^{(d,\boldsymbol{h};\boldsymbol{p})}(\boldsymbol{m})$.


\subsection{Scalar Five-Point Conformal Blocks in Comb Channel}

From \eqref{EqCB}, the scalar five-point conformal blocks in the comb channel are explicitly given by
\eqna{
&G_5^{(d,h_2,h_3,h_4;p_2,p_3,p_4)}(u_1^5,u_2^5,v_{11}^5,v_{12}^5,v_{22}^5)=\sum_{\{m_a,m_{ab}\}\geq0}(p_3)_{m_1+m_{11}+m_{22}}(p_4-m_1)_{m_2}\\
&\phantom{=}\qquad\times\frac{(p_2+h_2)_{m_1+m_{12}}(\bar{p}_3+\bar{h}_3)_{m_1+m_2+m_{11}+m_{12}+m_{22}}(\bar{p}_4+\bar{h}_4)_{m_2+m_{12}+m_{22}}}{(\bar{p}_3+h_2)_{2m_1+m_{11}+m_{12}+m_{22}}(\bar{p}_4+\bar{h}_3)_{2m_2+m_{11}+m_{12}+m_{22}}}\\
&\phantom{=}\qquad\times\frac{(-h_3)_{m_1}(-h_4)_{m_2}(-h_4+m_2)_{m_{11}}}{(\bar{p}_3+h_2+1-d/2)_{m_1}(\bar{p}_4+\bar{h}_3+1-d/2)_{m_2}}{}_3F_{2}\left[\begin{array}{c}-m_1,-m_2,-\bar{p}_3-h_2+d/2-m_1\\p_4-m_1,h_3+1-m_1\end{array};1\right]\\
&\phantom{=}\qquad\times\frac{(u_1^5)^{m_1}}{m_1!}\frac{(u_2^5)^{m_2}}{m_2!}\frac{(1-v_{11}^5)^{m_{11}}}{m_{11}!}\frac{(1-v_{12}^5)^{m_{12}}}{m_{12}!}\frac{(1-v_{22}^5)^{m_{22}}}{m_{22}!}.
}[EqCB5]
The result \eqref{EqCB5} is highly reminiscent of the result found in \cite{Rosenhaus:2018zqn}, but it is not the same.  It is possible to prove analytically that both results are equivalent using simple hypergeometric identities.

By expanding the ${}_3F_{2}$ as a sum
\eqn{{}_3F_{2}\left[\begin{array}{c}-m_1,-m_2,-\bar{p}_3-h_2+d/2-m_1\\p_4-m_1,h_3+1-m_1\end{array};1\right]=\sum_{n\geq0}\frac{(-m_1)_n(-m_2)_n(-\bar{p}_3-h_2+d/2-m_1)_n}{(p_4-m_1)_n(h_3+1-m_1)_nn!},}
and using the simple identities
\eqn{
\begin{gathered}
\frac{(-h_3)_{m_1-n}}{(\bar{p}_3+h_2+1-d/2)_{m_1-n}}=\frac{(-h_3)_{m_1}(-\bar{p}_3-h_2+d/2-m_1)_n}{(\bar{p}_3+h_2+1-d/2)_{m_1}(h_3+1-m_1)_n},\\
(p_4-m_1+n)_{m_2-n}=\frac{(p_4-m_1)_{m_2}}{(p_4-m_1)_n},\\
\frac{1}{(m_1-n)!(m_2-n)!}=\frac{(-m_1)_n(-m_2)_n}{m_1!m_2!},
\end{gathered}
}
the conformal block \eqref{EqCB5} can be re-expressed as
\eqna{
&G_5^{(d,h_2,h_3,h_4;p_2,p_3,p_4)}(u_1^5,u_2^5,v_{11}^5,v_{12}^5,v_{22}^5)=\sum_{\{m_a,m_{ab},n\}\geq0}(p_3)_{m_1+m_{11}+m_{22}}(p_4-m_1+n)_{m_2-n}\\
&\phantom{=}\qquad\times\frac{(p_2+h_2)_{m_1+m_{12}}(\bar{p}_3+\bar{h}_3)_{m_1+m_2+m_{11}+m_{12}+m_{22}}(\bar{p}_4+\bar{h}_4)_{m_2+m_{12}+m_{22}}}{(\bar{p}_3+h_2)_{2m_1+m_{11}+m_{12}+m_{22}}(\bar{p}_4+\bar{h}_3)_{2m_2+m_{11}+m_{12}+m_{22}}}\\
&\phantom{=}\qquad\times\frac{(-h_3)_{m_1-n}(-h_4)_{m_2+m_{11}}}{(\bar{p}_3+h_2+1-d/2)_{m_1-n}(\bar{p}_4+\bar{h}_3+1-d/2)_{m_2}}\\
&\phantom{=}\qquad\times\frac{(u_1^5)^{m_1}}{(m_1-n)!}\frac{(u_2^5)^{m_2}}{(m_2-n)!}\frac{(1-v_{11}^5)^{m_{11}}}{m_{11}!}\frac{(1-v_{12}^5)^{m_{12}}}{m_{12}!}\frac{(1-v_{22}^5)^{m_{22}}}{(m_{22}!)^2}.
}
Now we use the following identity,
\eqn{(p_4-m_1+n)_{m_2-n}=\sum_{j\geq0}\frac{(m_2-n)!}{j!(m_2-n-j)!}(p_4)_j(-m_1+n)_{m_2-n-j},}
with the change of summation variable $j\to m_2-j$, to get
\eqn{(p_4-m_1+n)_{m_2-n}=\sum_{j\geq0}\frac{(m_2-n)!(-j)_n}{(m_2-j)!j!}\frac{(p_4)_{m_2}}{(1-p_4-m_2)_j}\frac{(m_1-n)!}{(m_1-j)!},}
Since the only other terms in the conformal blocks including an explicit $n$ are
\eqn{\frac{(-h_3)_{m_1-n}}{(\bar{p}_3+h_2+1-d/2)_{m_1-n}}=\frac{(-h_3)_{m_1}(-\bar{p}_3-h_2+d/2)_n}{(\bar{p}_3+h_2+1-d/2)_{m_1}(h_3+1)_n},}
the sum over $n$ simplifies to
\eqn{\sum_n=\frac{(\bar{p}_3+\bar{h}_3+1-d/2)_j}{(h_3+1-m_1)_j}.}
Re-expressing the sum over $j$ as a ${}_3F_2$ leads to
\eqna{
&G_5^{(d,h_2,h_3,h_4;p_2,p_3,p_4)}(u_1^5,u_2^5,v_{11}^5,v_{12}^5,v_{22}^5)=\sum_{\{m_a,m_{ab}\}\geq0}(p_3)_{m_1+m_{11}+m_{22}}(p_4)_{m_2}\\
&\phantom{=}\qquad\times\frac{(p_2+h_2)_{m_1+m_{12}}(\bar{p}_3+\bar{h}_3)_{m_1+m_2+m_{11}+m_{12}+m_{22}}(\bar{p}_4+\bar{h}_4)_{m_2+m_{12}+m_{22}}}{(\bar{p}_3+h_2)_{2m_1+m_{11}+m_{12}+m_{22}}(\bar{p}_4+\bar{h}_3)_{2m_2+m_{11}+m_{12}+m_{22}}}\\
&\phantom{=}\qquad\times\frac{(-h_3)_{m_1}(-h_4)_{m_2+m_{11}}}{(\bar{p}_3+h_2+1-d/2)_{m_1}(\bar{p}_4+\bar{h}_3+1-d/2)_{m_2}}{}_3F_{2}\left[\begin{array}{c}-m_1,-m_2,\bar{p}_3+\bar{h}_3+1-d/2\\1-p_4-m_2,h_3+1-m_1\end{array};1\right]\\
&\phantom{=}\qquad\times\frac{(u_1^5)^{m_1}}{m_1!}\frac{(u_2^5)^{m_2}}{m_2!}\frac{(1-v_{11}^5)^{m_{11}}}{m_{11}!}\frac{(1-v_{12}^5)^{m_{12}}}{m_{12}!}\frac{(1-v_{22}^5)^{m_{22}}}{m_{22}!}.
}
which matches exactly with \cite{Rosenhaus:2018zqn} after the proper re-definitions $\eta_1\to\eta_5$, $\eta_a\to\eta_{a-1}$, $\Delta_1\to\Delta_5$, and $\Delta_a\to\Delta_{a-1}$ to match the operator positions, which imply
\eqn{u_1^5=u_1^R,\qquad u_2^5=u_2^R,\qquad v_{11}^5=v_1^R,\qquad v_{12}^5=v_2^R,\qquad v_{22}^5=w^R,}
where the cross-ratios with superscript $R$ are the ones defined in \cite{Rosenhaus:2018zqn}.


\subsection{Limit \texorpdfstring{$d\to1$}{d->1}}

It is also possible to verify the conformal blocks \eqref{EqCB} by comparing with the $d=1$ conformal blocks obtained in \cite{Rosenhaus:2018zqn}.

First, it is important to bring up that in general spacetime dimension, the number of cross-ratios for $M$-point correlation functions is usually stated as being $M(M-3)/2$.  However, due to the restrictions encountered for any fixed spacetime dimension, the number of independent cross-ratios is smaller than the usual value $M(M-3)/2$ when $M$ is large enough.  Indeed, for fixed spacetime dimension with large $M$, there are not enough spacetime dimensions to make all cross-ratios independent.  For example, in $d=3$, there are two independent cross-ratios for four-point conformal blocks, five independent cross-ratios for five-point conformal blocks, but only eight independent cross-ratios for six-point conformal blocks instead of the usual nine.  Generically, one has \cite{Ferrara:1974nf}
\eqn{
\begin{gathered}
N_{cr}=\frac{M(M-3)}{2},\qquad M-3<d,\\
N_{cr}=d(M-3)-\frac{(d-1)(d-2)}{2},\qquad M-3\geq d,
\end{gathered}
}[EqNcr]
for the number of independent cross-ratios $N_{cr}$.  The rationale behind \eqref{EqNcr} for $M-3\geq d$ is simple.  As usual, using conformal invariance three points are set at $0$, $1$ and $\infty$, fixing a line in the $d$-dimensional space.  There are thus $(M-3)$ remaining coordinates that can be placed anywhere in the $d$-dimensional space, up to the rotations about the fixed line, which corresponds to the term $(d-1)(d-2)/2$.

In one spacetime dimension, there are only $M-3$ independent cross-ratios for $M$-point conformal blocks.  In \cite{Rosenhaus:2018zqn}, those were defined as $\chi_a$ with $1\leq a\leq M-3$.  By comparing their explicit definitions with our definitions \eqref{EqCR}, it is easy to check that the cross-ratios used here are related to the cross-ratios $\chi_a$ as follows
\eqn{
\begin{gathered}
u_a^M\to\chi_a^2,\qquad1\leq a\leq M-3\\
v_{ab}^M\to\left(-1-\sum_{n=1}^{\lfloor\frac{a+1}{2}\rfloor}(-1)^n\sum_{c_1=1}^{a-2(n-1)}\sum_{c_2=c_1+2}^{a-2(n-2)}\cdots\sum_{c_n=c_{n-1}+2}^a\prod_{i=1}^n\chi_{c_i+b-a}\right)^2,\qquad1\leq a\leq b\leq M-3.
\end{gathered}
}[Eqd1]
Although it is not straightforward to check analytically, we did verify to some finite order in the cross-ratio expansions that the $(M\leq8)$-point conformal blocks \eqref{EqCB} with the substitutions \eqref{Eqd1} reproduced the $d=1$ conformal blocks obtained in \cite{Rosenhaus:2018zqn}.

Since the conformal blocks in $d=1$ obtained in \cite{Rosenhaus:2018zqn} are much simpler, it would be interesting to find an analytic proof of the equivalence with \eqref{EqCB}.  Such a proof could be useful in looking for simplifications to the higher-point conformal blocks in fixed spacetime dimensions when some cross-ratios are not independent.


\section{Discussion and Conclusion}\label{SecConc}

Our main results are given in \eqref{EqCB} and \eqref{EqF}.  These formulas encode the $M$-point scalar conformal blocks in the comb channel for any dimension $d$.  This result was obtained using the embedding space OPE approach for computing conformal blocks set out in \cite{Fortin:2019fvx,Fortin:2019dnq}.  This result passes several consistency checks.  In appropriate limits, it agrees with the results of \cite{Rosenhaus:2018zqn}.  Furthermore, when one of the external operators is exchanged for the identity operator our expressions for an $M$-point block reduces to that for an $(M-1)$-point block.

There are several directions in which it might be interesting to extended these results.  Starting with $M=6$, there are other configurations beyond the comb, in which some of the OPE vertices are not directly connected to any external operators.  The OPE formalism is not limited to the comb channel, but can be applied to other cases as well.  It would be interesting to see relations between different topologies of the blocks if progress can be made in computing blocks of different topologies efficiently.  Likewise, it is feasible to compute higher-point functions containing exchange operators with spin.  At least some of such cases do not appear to be prohibitively complicated, and in any case seem simpler than higher-point functions with spinning external operators.

The $M$-point blocks described in \eqref{EqCB} and \eqref{EqF} are described by a complicated function of the invariant cross-ratios.  This function must have many remarkable properties that would be worth studying.  It is not clear if there are other choices of the invariant cross-ratios that may make some of the properties of the blocks more apparent.  For example, in the limit where one of the exchange operators has dimension set to $0$, neighboring operators must be forced to have identical dimensions.

While for the four-point blocks the AdS/CFT correspondence has been used fruitfully, there could be interesting applications of higher-point blocks.  The connection between conformal blocks on the boundary and the ``geodesic Witten diagrams" has been established in \cite{Hijano:2015zsa}.

Finally, knowledge of higher-point functions could be used in the bootstrap program.  The unitarity constraints that have led to many powerful results delineating the space of consistent CFTs, thus far, have been based on the positivity of the two-point functions.  Obviously, there are additional unitarity constraints that could be imposed and leveraged to further constrain CFTs.  It has also been suggested that studying $M$-point blocks with external scalar fields could be an alternative for studying crossing equations containing external operators with spin \cite{Rosenhaus:2018zqn}.

\vspace{0.5cm}
\noindent\textbf{Note added:} During completion of this work a result for $M$-point blocks in the comb channel appeared on the preprint archive \cite{Parikh:2019dvm}.  That result was obtained using holographic methods generalizing lower-point results by conjecture and subsequently verified (up to $M=7$) using the Casimir equation.


\ack{
The authors would like to thank Valentina Prilepina for useful discussions.  The work of JFF is supported by NSERC and FRQNT.  The work of WJM is supported by the China Scholarship Council and in part by NSERC and FRQNT.
}


\setcounter{section}{0}
\renewcommand{\thesection}{\Alph{section}}

\section{Notation and the Differential Operator}\label{SAppNotation}

We denote the dimensionality of spacetime by $d$, but work in the embedding space which is a $(d+2)$-dimensional projective space.  The embedding space coordinates, denoted $\eta$, are restricted to the light-cone $\eta\cdot\eta\equiv\eta_A\eta^A=0$ and equivalent under $\eta^A\sim\lambda\eta^A$ for $\lambda>0$.  The conformal symmetry acts linearly on the embedding space coordinates $\eta^A$.  Since we are dealing with scalar primary operators all expressions depend on the dot products alone that is on $\eta_i\cdot\eta_j=-\frac{1}{2}(x_i-x_j)^2$, where $x_i$ is the ordinary $d$-dimensional position space coordinate corresponding to $\eta_i$.

The primary operators $\Op{}{}$ in the embedding space are homogenous under coordinate re-scalings $\eta^A\frac{\partial}{\partial\eta^A}\Op{}{} =-\Delta_{\mathcal{O}}\Op{}{}$, where $\Delta_{\mathcal{O}}$ is the operator dimension.  Operators in non-trivial Lorentz representations satisfy additional transversality conditions and scale according to their twists instead of their dimensions \cite{Fortin:2019dnq}, but these intricacies are of no consequence here.

We now turn to the differential operator in the OPE for scalar operators \eqref{EqOPE-scalar}
\eqn{
\begin{gathered}
\Op{i}{1}\Op{j}{2}=\cOPE{}{i}{j}{k}\frac{1}{\ee{1}{2}{p_{ijk}}}\D_{12}^{(d,h_{ijk},0)}\Op{k}{2}+\ldots,\\
p_{ijk}=\frac{1}{2}(\Delta_i+\Delta_j-\Delta_k),\qquad h_{ijk}=-\frac{1}{2}(\Delta_i-\Delta_j+\Delta_k).
\end{gathered}}
The differential operator $\D_{12}^{(d,h_{ijk},0)}$ is rather simple when its third superscript, corresponding to the number of Lorentz indices, is zero.  This differential operator can be defined for any pairs of coordinates
\eqn{\D_{ij}^{(d,h,0)}=\left[\ee{i}{j}{}\partial_j^2-(d+2\,\eta_j\cdot\partial_j)\eta_i\cdot\partial_j\right]^h,}
where $\partial_j=\frac{\partial}{\partial\eta_j^A}$.  Note that $\D_{ij}^{(d,h,0)}$ contains derivatives with respect to $\eta_j$ only, but depends on both coordinates $\eta_i$ and $\eta_j$.  The power $h$ is, in general, not necessarily integer.  However, the action of $\D_{ij}^{(d,h,0)}$ on the coordinates can be defined for any real $h$ using fractional calculus \cite{Fortin:2016dlj,Comeau:2019xco}.

$\D_{ij}^{(d,h,0)}$ satisfies a number of useful properties, for example $\D_{ij}^{(d,h,0)}\ee{i}{j}{}=\ee{i}{j}{}\D_{ij}^{(d,h,0)}$.  $\D_{ij}^{(d,h,0)}$ is homogeneous of degree $h$ with respect to $\eta_i$, that is $[\eta_i\cdot\partial_i,\D_{ij}^{(d,h,0)}]=h\D_{ij}^{(d,h,0)}$, and of degree $-h$ with respect to $\eta_j$, that is $[\eta_j\cdot\partial_j,\D_{ij}^{(d,h,0)}]=-h \D_{ij}^{(d,h,0)}$.  What is crucial for us is that the action of the differential operator can be calculated explicitly in all generality.  The function $I_{ij}^{(d,h,0,\boldsymbol{p})}$ defined as
\eqn{I_{ij}^{(d,h,0;\boldsymbol{p})}=\D_{ij}^{(d,h,0)}\prod_{a\neq i,j}\frac{1}{\ee{j}{a}{p_a}},}[EqAppI]
where $\boldsymbol{p}$ denotes collectively all powers $p_a$, is the most general expression one encounters when computing the OPE of scalar operators, that is when an $M$-point function is expressed in terms of an $(M-1)$-point function.

With four, or more, coordinates it is possible to introduce a differential operator that is homogenous of degree $0$ with respect to every coordinate
\eqn{\bar{\D}_{ij;k\ell}^{(d,h,0)}=\frac{\ee{i}{j}{h}\ee{k}{\ell}{h}}{\ee{i}{k}{h}\ee{i}{\ell}{h}} \D_{ij}^{(d,h,0)},}
where $k,\ell\neq i,j$.  All the dot products in the pre-factor in the definition of $\bar{\D}_{ij;k\ell}^{(d,h,0)}$ commute with the differential operator.  Similarly, it is possible to introduce a re-scaled homogeneous analogue of \eqref{EqAppI}
\eqn{\bar{I}_{ij;k\ell}^{(d,h,0;\boldsymbol{p})}=\frac{\ee{i}{j}{\bar{p}+h}\ee{k}{\ell}{\bar{p}+h}}{\ee{i}{k}{\bar{p}+h}\ee{i}{\ell}{\bar{p}+h}}\left[\prod_{a\neq i,j}\ee{i}{a}{p_a}\right]I_{ij}^{(d,h,0;\boldsymbol{p})}=\bar{\D}_{ij;k\ell}^{(d,h,0)}\prod_{a\neq i,j}x_a^{p_a},}[EqIb]
with $\bar{p}=\sum_{a\neq i,j}p_a$.  Because $\bar{I}_{ij;k\ell}^{(d,h,0;\boldsymbol{p})}$ is homogeneous of degree $0$ in all coordinates it can be described as a function of conformal cross-ratios $x_a=\frac{\ee{i}{j}{}\ee{k}{\ell}{}\ee{i}{a}{}}{\ee{i}{k}{}\ee{i}{\ell}{}\ee{j}{a}{}}$ for all $a\neq i,j$, as indicated in the second equality in \eqref{EqIb}.  For computational convenience, it is useful to express the differential operators not in terms of the cross-ratios $x_a$, but instead to pick one specific $x_m$ and trade the remaining $x_a$'s for $y_a=1-x_m/x_a$ for $a\neq i,j,m$.  When the operator $\bar{\D}_{ij;k\ell}^{(d,h,0)}$ is expressed in terms of $x_m$ and $y_a$, by a tedious yet straightforward change of variables, it is referred to as $\bar{\D}_{ij;k\ell;m}^{2h}$.  Further details are in Section 4.3 of \cite{Fortin:2019dnq}.  The action of $\bar{\D}_{ij;k\ell;m}^{2h}$ used in this article differs, for brevity of expressions, by a purely numerical factor from that in \cite{Fortin:2019dnq}.  In \eqref{EqD} a factor of  $(-2)^{h_{M-1}}(\bar{q})_{h_{M-1}}(\bar{q}+1-d/2)_{h_{M-1}}$ is omitted compared to the analogous expression in \cite{Fortin:2019dnq}.  Since the difference is purely numerical it does not affect coordinate dependence.


\section{Sketch of the Recurrence Relation \eqref{EqRR}}\label{SAppOPE}

As mentioned in the main text, the recurrence relation for $F_M^{(d,\boldsymbol{h};\boldsymbol{p})}(\boldsymbol{m})$ given by \eqref{EqRR} can be obtained directly from the definition \eqref{EqGG}, itself obtained from the OPE \eqref{EqOPE}.  Since the action of the OPE differential operator on the cross-ratios \eqref{EqCROPE} is known, using \eqref{Equtox} one first expresses the cross-ratios \eqref{EqCR} in terms of the cross-ratios \eqref{EqCROPE}.  Then, the action of the OPE differential operator is computed from \eqref{EqD}, and the result is re-expressed in terms of the cross-ratios \eqref{EqCR} using \eqref{Eqxtou}.  This leads to
\eqna{
&G_M^{(d,\boldsymbol{h};\boldsymbol{p})}(\boldsymbol{u}^M,\textbf{v}^M)\\
&\qquad=\sum\prod_{i=1}^{M-4}\frac{1}{(n_{i,M-4})!}\binom{n_{i,M-4}}{q_{M-i-2}}\binom{-\bar{p}_{M-1}-\bar{h}_{M-1}-m_{M-3}-\sum_{a=3}^{M-2}l_a}{m_{M-3,M-3}}\binom{q_2-\bar{r}_2}{m_{M-4,M-4}}\\
&\qquad\phantom{=}\times\prod_{a=3}^{M-2}\binom{l_a}{m_{M-a-1,M-3}}\prod_{a=3}^{M-3}\binom{q_a-l_a-s_a}{m_{M-a-2,M-4}}\prod_{a=1}^{M-5}\prod_{b=1}^a\binom{r_{a-b+2,a+4}}{k_{ab}}\prod_{a=3}^{M-1}(-1)^{l_a}\binom{\sigma_a}{l_a}\\
&\qquad\phantom{=}\times(-1)^{\sum_{a=2}^{M-3}q_a+\sum_{a=1}^{M-3}m_{a,M-3}+\sum_{a=1}^{M-4}m_{a,M-4}+\sum_{a=1}^{M-5}\sum_{b=1}^ak_{ab}}\\
&\qquad\phantom{=}\times\frac{(-h_{M-1})_{m_{M-3}}(-q_2)_{\bar{r}_2}(\bar{p}_{M-1}+\bar{h}_{M-1})_{m_{M-3}+\sum_{a=3}^{M-1}\sigma_a}}{(\bar{p}_{M-1}+\bar{h}_{M-2})_{2m_{M-3}+\sum_{a=3}^{M-1}\sigma_a}(\bar{p}_{M-1}+\bar{h}_{M-2}+1-d/2)_{m_{M-3}}}\\
&\qquad\phantom{=}\times\frac{(p_{M-1}-n_{M-4})_{\sigma_{M-1}+s_{M-1}}}{(\sigma_{M-1})!}\frac{(\bar{p}_{M-2}+\bar{h}_{M-2}+n_{M-4}+\sum_{a=2}^{M-3}q_a)_{\sigma_{M-2}+s_{M-2}}}{(\sigma_{M-2})!}\\
&\qquad\phantom{=}\times\prod_{a=3}^{M-3}\frac{(-q_a)_{\sigma_a+s_a}}{(\sigma_a)!}\prod_{\substack{b>a\\a,b\neq 1,M}}\frac{1}{(r_{ab})!}(u^M_{M-3})^{m_{M-2}}\prod_{i=1}^{M-4}\frac{(u^M_{i})^{n_i+r_{23}+\sum_{b=4}^{i+2}\sum_{a=2}^{b-1}r_{ab}}}{(n_i)!}\\
&\qquad\phantom{=}\times(1-v^M_{M-4,M-4})^{m_{M-4,M-4}}\prod_{a=1}^{M-4}(1-v^M_{a,M-3})^{m_{a,M-3}}\prod_{a=1}^{M-5}(1-v^M_{a,M-4})^{m_{a,M-4}}\\
&\qquad\phantom{=}\times\prod_{b=1}^{M-5}\prod_{a=1}^b\frac{(1-v^M_{ab})^{n_{ab}+k_{ab}}}{(n_{ab})!}\left(\frac{v^M_{M-3,M-3}}{v^M_{M-4,M-4}}\right)^{m_{M-3,M-3}}C_{M-1}^{(d,\boldsymbol{h};\boldsymbol{p})}(\boldsymbol{m},\textbf{m})F_{M-1}^{(d,\boldsymbol{h};\boldsymbol{p})}(\boldsymbol{m}),
}
where
\eqn{\bar{r}_2=\sum_{3\leq a\leq M-1}r_{2a},}
and $C_M^{(d,\boldsymbol{h};\boldsymbol{p})}(\boldsymbol{m},\textbf{m})$ is defined through \eqref{EqCB} as
\eqn{G_M^{(d,\boldsymbol{h};\boldsymbol{p})}(\boldsymbol{u}^M,\textbf{v}^M)=\sum_{\{m_a,m_{ab}\}\geq0}C_M^{(d,\boldsymbol{h};\boldsymbol{p})}(\boldsymbol{m},\textbf{m})F_M^{(d,\boldsymbol{h};\boldsymbol{p})}(\boldsymbol{m})\prod_{1\leq a\leq M-3}\frac{(u_a^M)^{m_a}}{m_a!}\prod_{\substack{a,b\\b\geq a}}\frac{(1-v_{ab}^M)^{m_{ab}}}{m_{ab}!}.}

We are thus left with sums over the usual $m_a$ and $m_{ab}$ as well as extra sums over $n_{ab}$, $q_a$, $l_a$, $r_{ab}$, $s_a$, $k_{ab}$, and $\sigma_a$ that we want to re-sum.  By proceeding appropriately, all these sums are of the Gauss' hypergeometric type ${}_2F_1(-m,b;c;1)=\frac{(c-b)_m}{(c)_m}$ for $m$ a non-negative integer.  For example, for the sum over $l_{M-1}$, the relevant terms are
\eqn{(-1)^{l_{M-1}}\binom{\sigma_{M-1}}{l_{M-1}}=\frac{(-\sigma_{M-1})_{l_{M-1}}}{l_{M-1}!},}
and
\eqna{
&(-1)^{m_{M-3,M-3}}\binom{\bar{p}_{M-1}+\bar{h}_{M-1}+m_{M-3}+\sum_{a=3}^{M-1}l_a}{m_{M-3,M-3}}\\
&\qquad=\frac{(\bar{p}_{M-1}+\bar{h}_{M-1}+m_{M-3}+\sum_{a=3}^{M-2}l_a+m_{M-3,M-3})_{l_{M-1}}(\bar{p}_{M-1}+\bar{h}_{M-1}+m_{M-3}+\sum_{a=3}^{M-2}l_a)_{m_{M-3,M-3}}}{(\bar{p}_{M-1}+\bar{h}_{M-1}+m_{M-3}+\sum_{a=3}^{M-2}l_a)_{l_{M-1}}m_{M-3,M-3}!}.
}
Therefore, the sum over $l_{M-1}$ is simply
\eqn{\frac{(-m_{M-3,M-3})_{\sigma_{M-1}}(\bar{p}_{M-1}+\bar{h}_{M-1}+m_{M-3}+\sum_{a=3}^{M-2}l_a)_{m_{M-3,M-3}}}{(\bar{p}_{M-1}+\bar{h}_{M-1}+m_{M-2}+\sum_{a=3}^{M-2}l_a)_{\sigma_{M-1}}m_{M-3,M-3}!}.}
To get to \eqref{EqRR}, we repeat the steps above for all remaining indices of summation, which is long yet straightforward.


\section{Proof of \eqref{EqF} from the Recurrence Relation \eqref{EqRR}}\label{SAppRR}

The recurrence relation \eqref{EqRR} directly leads to a proof of \eqref{EqF}.  Indeed, one has
\eqna{
&F_M^{(d,\boldsymbol{h};\boldsymbol{p})}(\boldsymbol{m})\\
&=\sum_{\{t_a\}\geq0}(-m_{M-3})_{t_{M-4}}\prod_{1\leq a\leq M-4}\frac{(-m_a)_{t_a}(-\bar{p}_{a+2}-\bar{h}_{a+1}+d/2-m_a)_{t_a}(-t_a)_{t_{a-1}}}{(p_{a+3}-m_a)_{t_a}(h_{a+2}+1-m_a)_{t_a}t_a!}F_{M-1}^{(d,\boldsymbol{h};\boldsymbol{p})}(\boldsymbol{m}-\boldsymbol{t})\\
&=\sum_{\{t_a\}\geq0}(-m_{M-3})_{t_{M-4}}\prod_{1\leq a\leq M-4}\frac{(-m_a)_{t_a}(-\bar{p}_{a+2}-\bar{h}_{a+1}+d/2-m_a)_{t_a}(-t_a)_{t_{a-1}}}{(p_{a+3}-m_a)_{t_a}(h_{a+2}+1-m_a)_{t_a}t_a!}\\
&\phantom{=}\qquad\times\prod_{1\leq a\leq M-5}{}_3F_2\left[\begin{array}{c}-m_a+t_a,-m_{a+1}+t_{a+1},-\bar{p}_{a+2}-\bar{h}_{a+1}+d/2-m_a+t_a\\p_{a+3}-m_a+t_a,h_{a+2}+1-m_a+t_a\end{array};1\right]\\
&=\sum_{\{t_a\}\geq0}(-m_{M-3})_{t_{M-4}}\prod_{1\leq a\leq M-4}\frac{(-m_a)_{t_a}(-\bar{p}_{a+2}-\bar{h}_{a+1}+d/2-m_a)_{t_a}(-t_a)_{t_{a-1}}}{(p_{a+3}-m_a)_{t_a}(h_{a+2}+1-m_a)_{t_a}t_a!}\\
&\phantom{=}\qquad\times\sum_{\{j_a\}\geq0}\prod_{a=1}^{M-5}\frac{(-m_a+t_a)_{j_a}(-m_{a+1}+t_{a+1})_{j_a}(-\bar{p}_{a+2}-\bar{h}_{a+1}+d/2-m_a+t_a)_{j_a}}{(p_{a+3}-m_a+t_a)_{j_a}(h_{a+2}+1-m_a+t_a)_{j_a}j_a!}\\
&=\sum_{\{t_a\}\geq0}(-m_{M-3})_{t_{M-4}}\frac{(-m_{M-4})_{t_{M-4}}(-\bar{p}_{M-2}-\bar{h}_{M-3}+d/2-m_{M-4})_{t_{M-4}}}{(p_{M-1}-m_{M-4})_{t_{M-4}}(h_{M-2}+1-m_{M-4})_{t_{M-4}}t_{M-4}!}\\
&\phantom{=}\qquad\times\sum_{\{j_a\}\geq0}\prod_{a=1}^{M-5}\frac{(-m_a)_{j_a+t_a}(-m_{a+1}+t_{a+1})_{j_a}(-\bar{p}_{a+2}-\bar{h}_{a+1}+d/2-m_a)_{j_a+t_a}(-t_{a+1})_{t_a}}{(p_{a+3}-m_a)_{j_a+t_a}(h_{a+2}+1-m_a)_{j_a+t_a}j_a!t_a!}\\
&=\sum_{\{j_a\}\geq0}(-m_{M-3})_{j_{M-4}}\frac{(-m_{M-4})_{j_{M-4}}(-\bar{p}_{M-2}-\bar{h}_{M-3}+d/2-m_{M-4})_{j_{M-4}}}{(p_{M-1}-m_{M-4})_{j_{M-4}}(h_{M-2}+1-m_{M-4})_{j_{M-4}}j_{M-4}!}\\
&\phantom{=}\qquad\times\sum_{\{t_a\}\geq0}\prod_{a=1}^{M-5}\frac{(-m_a)_{j_a}(-m_{a+1}+t_{a+1})_{j_a-t_a}(-\bar{p}_{a+2}-\bar{h}_{a+1}+d/2-m_a)_{j_a}(-t_{a+1})_{t_a}}{(p_{a+3}-m_a)_{j_a}(h_{a+2}+1-m_a)_{j_a}(j_a-t_a)!t_a!},
}
where in the last line we changed the summation variables from $j_a\to j_a-t_a$ for $1\leq a\leq M-5$ and renamed $t_{M-4}$ by $j_{M-4}$.  Combining the products, the result can be re-expressed as
\eqna{
F_M^{(d,\boldsymbol{h};\boldsymbol{p})}(\boldsymbol{m})&=\sum_{\{j_a\}\geq0}\frac{(-m_{M-3})_{j_{M-4}}}{j_{M-4}!}\prod_{a=1}^{M-4}\frac{(-m_a)_{j_a}(-\bar{p}_{a+2}-\bar{h}_{a+1}+d/2-m_a)_{j_a}}{(p_{a+3}-m_a)_{j_a}(h_{a+2}+1-m_a)_{j_a}}\\
&\phantom{=}\qquad\times\sum_{\{t_a\}\geq0}\prod_{a=1}^{M-5}\frac{(-m_{a+1}+t_{a+1})_{j_a-t_a}(-t_{a+1})_{t_a}}{(j_a-t_a)!t_a!},
}
where it is understood that $t_{M-4}=j_{M-4}$.

Concentrating on the $t_1$ sum and using simple hypergeometric identities of the Gauss' type, we get
\eqna{
\sum_{t_1\geq0}\frac{(-m_2+t_2)_{j_1-t_1}(-t_2)_{t_1}}{(j_1-t_1)!t_1!}=\frac{(-m_2)_{j_1}}{j_1!},
}
which is independent of $t_2$, leading to
\eqna{
F_M^{(d,\boldsymbol{h};\boldsymbol{p})}(\boldsymbol{m})&=\sum_{\{j_a\}\geq0}\frac{(-m_2)_{j_1}}{j_1!}\frac{(-m_{M-3})_{j_{M-4}}}{j_{M-4}!}\prod_{a=1}^{M-4}\frac{(-m_a)_{j_a}(-\bar{p}_{a+2}-\bar{h}_{a+1}+d/2-m_a)_{j_a}}{(p_{a+3}-m_a)_{j_a}(h_{a+2}+1-m_a)_{j_a}}\\
&\phantom{=}\qquad\times\sum_{\{t_a\}\geq0}\prod_{a=2}^{M-5}\frac{(-m_{a+1}+t_{a+1})_{j_a-t_a}(-t_{a+1})_{t_a}}{(j_a-t_a)!t_a!},
}
Clearly, because of its independence on $t_{a+1}$ the same identity can be used recursively on the $t_a$ sums, starting from $a=1$ all the way to $a=M-5$, giving
\eqn{F_M^{(d,\boldsymbol{h};\boldsymbol{p})}(\boldsymbol{m})=\sum_{\{j_a\}\geq0}\prod_{a=1}^{M-4}\frac{(-m_a)_{j_a}(-m_{a+1})_{j_a}(-\bar{p}_{a+2}-\bar{h}_{a+1}+d/2-m_a)_{j_a}}{(p_{a+3}-m_a)_{j_a}(h_{a+2}+1-m_a)_{j_a}j_a!},}
which is nothing else than \eqref{EqF}, completing the proof.


\section{Sketch of the Proof of the Limit of Unit Operator}\label{SAppUnit}

The limit of unit operator $\Op{i_{M-1}}{M-1}\to\mathds{1}$ implies that $p_{M-1}=h_{M-2}=0$.  Through the Pochhammer symbols, these constraints force some of the summation variables to vanish.  Moreover, the cross-ratios transform such that
\eqn{
\begin{gathered}
\left.u_{M-5}^{M-1}\right|_{\eta_{M-1}\to\eta_M}=\frac{u_{M-5}^M}{v_{1,M-4}^M},\\
\left.u_{M-4}^{M-1}\right|_{\eta_{M-1}\to\eta_M}=\frac{u_{M-4}^Mu_{M-3}^M}{v_{1,M-4}^Mv_{1,M-3}^M},\\
\left.v_{a,M-5}^{M-1}\right|_{\eta_{M-1}\to\eta_M}=\frac{v_{a+1,M-4}^M}{v_{1,M-4}^M},\qquad1\leq a\leq M-5,\\
\left.v_{a,M-4}^{M-1}\right|_{\eta_{M-1}\to\eta_M}=\frac{v_{a+1,M-3}^M}{v_{1,M-4}^Mv_{1,M-3}^M},\qquad1\leq a\leq M-4.
\end{gathered}
}
Re-writing the conformal blocks in terms of the cross-ratios for $M-1$ points and re-summing with the help of several changes of variables lead to
\eqna{
G_M^{(d,\boldsymbol{h};\boldsymbol{p})}(\boldsymbol{u}^M,\textbf{v}^M)&\to(v_{1,M-4}^M)^{-\bar{p}_{M-3}-\bar{h}_{M-3}}(v_{1,M-3}^M)^{-\bar{p}_{M-1}-\bar{h}_{M-1}}\\
&\phantom{\to}\qquad\times G_{M-1}^{(d,\boldsymbol{h};\boldsymbol{p})}\left(\left.\boldsymbol{u}^{M-1}\right|_{\eta_{M-1}\to\eta_M},\left.\textbf{v}^{M-1}\right|_{\eta_{M-1}\to\eta_M}\right),
}
when $\Op{i_{M-1}}{M-1}\to\mathds{1}$.

Hence the contributions to the correlation function \eqref{EqCF} transform as
\eqna{
I_{M(\Delta_{k_1},\ldots,\Delta_{k_{M-3}})}^{(\Delta_{i_2},\ldots,\Delta_{i_M},\Delta_{i_1})}&\to(v_{1,M-4}^M)^{-\Delta_{i_{M-2}}/2}(v_{1,M-3}^M)^{(\Delta_{i_1}-\Delta_{i_M})/2}\\
&\phantom{\to}\qquad\times(v_{1,M-4}^M)^{(\Delta_{k_{M-5}}+\Delta_{k_{M-4}})/2}(v_{1,M-3}^M)^{\Delta_{k_{M-4}}/2}\\
&\phantom{\to}\qquad\times(v_{1,M-4}^M)^{-\bar{p}_{M-3}-\bar{h}_{M-3}}(v_{1,M-3}^M)^{-\bar{p}_{M-1}-\bar{h}_{M-1}}I_{M-1(\Delta_{k_1},\ldots,\Delta_{k_{M-4}})}^{(\Delta_{i_2},\ldots,\Delta_{i_{M-2}},\Delta_{i_M},\Delta_{i_1})}\\
&=I_{M-1(\Delta_{k_1},\ldots,\Delta_{k_{M-4}})}^{(\Delta_{i_2},\ldots,\Delta_{i_{M-2}},\Delta_{i_M},\Delta_{i_1})},
}
when $\Op{i_{M-1}}{M-1}\to\mathds{1}$ since
\eqn{\bar{p}_a+\bar{h}_{a-1}=\Delta_{k_{a-2}},\qquad2\leq a\leq M-1.}
Here, the pre-factor $(v_{1,M-4}^M)^{-\Delta_{i_{M-2}}/2}(v_{1,M-3}^M)^{(\Delta_{i_1}-\Delta_{i_M})/2}$ comes from the $\eta$'s in \eqref{EqCF} while the pre-factor $(v_{1,M-4}^M)^{(\Delta_{k_{M-5}}+\Delta_{k_{M-4}})/2}(v_{1,M-3}^M)^{\Delta_{k_{M-4}}/2}$ comes from the $u$'s in \eqref{EqCF} when $\Op{i_{M-1}}{M-1}\to\mathds{1}$.  Therefore the contribution to the correlation functions behaves properly under the limit of unit operator.


\subsection{Proof of the Limit of Unit Operator for Five-Point Functions}

To see how the mathematics work, we show here the complete proof of the limit of unit operator for five-point functions.

As argued above, in the limit $\mathcal{O}_{i_4}(\eta_4)\rightarrow\mathds{1}$, we should have
\eqn{I_{5(\Delta_{k_1},\Delta_{k_2})}^{(\Delta_{i_2},\ldots,\Delta_{i_5},\Delta_{i_1})}\to I_{4(\Delta_{k_1})}^{(\Delta_{i_2},\Delta_{i_3},\Delta_{i_5},\Delta_{i_1})}.}
To confirm this behavior, we first note that when $\Delta_4=0$ and $\Delta_{k_1}=\Delta_{k_2}$, we have\footnote{In the following, we use primes to label quantities with respect to $I_{4(\Delta_{k_1})}^{(\Delta_{i_2},\Delta_{i_3},\Delta_{i_5},\Delta_{i_1})}$.  We thus leave implicit most subscripts and superscripts to simplify the notation.}
\eqn{
\begin{gathered}
p_2^{\prime}=p_2,\qquad p_3^{\prime}=p_3\qquad p^{\prime}_4=p_5,\\
h^{\prime}_2=h_2,\qquad h^{\prime}_3=h_4.
\end{gathered}
}
Thus it is easy to check that under this limit, we have
\eqn{L_{5}(u^{5}_1)^{\Delta_{k_1}/2}(u^5_2)^{\Delta_{k_2}/2}\rightarrow(v^5_{11})^{p_3}(v^5_{12})^{\bar{p}_4+\bar{h}_4}L^{\prime}_4(u^{\prime 4}_1)^{\Delta_{k_1}/2},}[EqPrefactor]
where $L_M$ is the pre-factor with explicit $\eta$'s in \eqref{EqCF}, \textit{i.e.}\ the first line of \eqref{EqCF}.  To find the limiting behavior of $G_5$, we note that our definition gives
\eqna{
G_5&=\sum\frac{(p_3)_{m_1+m_{11}+m_{22}}(-h_3)_{m_1}(p_2+h_2)_{m_1+m_{12}}(\bar{p}_3+\bar{h}_3)_{m_1+m_2+m_{11}+m_{12}+m_{22}}}{(\bar{p}_3+h_2)_{2m_1+m_{11}+m_{12}+m_{22}}(\bar{p}_3+h_2+1-d/2)_{m_1}}\\
&\phantom{=}\qquad\times\frac{(p_4-m_1)_{m_2}(-h_4+m_2)_{m_{11}}(-h_4)_{m_2}(\bar{p}_4+\bar{h}_4)_{m_2+m_{12}+m_{22}}}{(\bar{p}_4+\bar{h}_3)_{2m_2+m_{11}+m_{12}+m_{22}}(\bar{p}_4+\bar{h}_3+1-d/2)_{m_2}}\\
&\phantom{=}\qquad\times\frac{(u^5_1)^{m_1}}{m_1!}\frac{(u^5_2)^{m_2}}{m_2!}\frac{(1-v^5_{11})^{m_{11}}}{m_{11}!}\frac{(1-v^5_{12})^{m_{12}}}{m_{12}!}\frac{(1-v^5_{22})^{m_{22}}}{m_{22}!}\\
&\phantom{=}\qquad\times\frac{(-m_1)_t(-m_2)_t(-\bar{p}_3-h_2+d/2-m_1)_t}{t!(p_4-m_1)_t(1+h_3-m_1)_t},
}
where we expanded $F_5$ as a summation over $t$, with all indices of summation left implicit in the sum.  Since in the limit above we have $p_4=h_3=0$, which implies $m_1=m_2=t$,
$G_5$ simplifies to
\eqna{
G_5&=\sum\frac{(p_3)_{m_1+m_{11}+m_{22}}(p_2+h_2)_{m_1+m_{12}}(-h_4+m_1)_{m_{11}}(-h_4)_{m_1}(\bar{p}_4+\bar{h}_4)_{m_1+m_{12}+m_{22}}}{(\bar{p}_4+\bar{h}_3)_{2m_1+m_{11}+m_{12}+m_{22}}(\bar{p}_4+\bar{h}_3+1-d/2)_{m_1}}\\
&\phantom{=}\qquad\times\frac{(u^5_1u^5_2)^{m_1}}{m_1!}\frac{(1-v^5_{11})^{m_{11}}}{m_{11}!}\frac{(1-v^5_{12})^{m_{12}}}{m_{12}!}\frac{(1-v^5_{22})^{m_{22}}}{m_{22}!}.
}
Using the fact that
\eqn{u^{\prime 4}_1=\frac{u^5_1u^5_2}{v^5_{11}v^5_{12}},\qquad v^{\prime 4}_{11}=\frac{v^5_{22}}{v^5_{11}v^5_{12}},}
we can rewrite $G_5$ in terms of $u^4_1$ and $v^4_{11}$.  This leads to
\eqna{
G_5&=\sum\frac{(p_3)_{m_1+m_{11}+m_{22}}(p_2+h_2)_{m_1+m_{12}}(-h_4+m_1)_{m_{11}}(-h_4)_{m_1}(\bar{p}_4+\bar{h}_4)_{m_1+m_{12}+m_{22}}}{(\bar{p}_4+\bar{h}_3)_{2m_1+m_{11}+m_{12}+m_{22}}(\bar{p}_4+\bar{h}_3+1-d/2)_{m_1}}\\
&\phantom{=}\qquad\times\binom{m_{22}}{k_{22}}\binom{k_{22}}{m^{\prime}_{11}}(-1)^{k_{22}+m^{\prime}_{11}}\frac{(u^{\prime 4}_1v^5_{11}v^5_{12})^{m_1}}{m_1!}\frac{(1-v^5_{11})^{m_{11}}}{m_{11}!}\frac{(1-v^5_{12})^{m_{12}}}{m_{12}!}\frac{(1-v^{\prime 4}_{11})^{m^{\prime}_{11}}(v^5_{11}v^5_{12})^{k_{22}}}{m_{22}!}.
}
Now, we change the variable to $k_{22}\rightarrow k_{22}+m^{\prime}_{11}$ such that
\eqn{\frac{(-1)^{k_{22}+m^{\prime}_{11}}}{m_{22}!}\binom{m_{22}}{k_{22}}\binom{k_{22}}{m^{\prime}_{11}}\rightarrow\frac{(-1)^{k_{22}}}{m_{22}!}\binom{m_{22}}{k_{22}+m^{\prime}_{11}}\binom{k_{22}+m^{\prime}_{11}}{m^{\prime}_{11}}=\frac{(-1)^{k_{22}}}{(m_{22}-m^{\prime}_{11})!m^{\prime}_{11}!}\binom{m_{22}-m^{\prime}_{11}}{k_{22}}.}
Therefore, $G_5$ becomes
\eqna{
G_5&=\sum\frac{(p_3)_{m_1+m_{11}+m_{22}}(p_2+h_2)_{m_1+m_{12}}(-h_4+m_1)_{m_{11}}(-h_4)_{m_1}(\bar{p}_4+\bar{h}_4)_{m_1+m_{12}+m_{22}}}{(\bar{p}_4+\bar{h}_3)_{2m_1+m_{11}+m_{12}+m_{22}}(\bar{p}_4+\bar{h}_3+1-d/2)_{m_1}}\\
&\phantom{=}\qquad\times\binom{m_{22}-m^{\prime}_{11}}{k_{22}}(-1)^{k_{22}}\frac{(u^{\prime 4}_1v^5_{11}v^5_{12})^{m_1}}{m_1!}\frac{(1-v^5_{11})^{m_{11}}}{m_{11}!}\frac{(1-v^5_{12})^{m_{12}}}{m_{12}!}\frac{(1-v^{\prime 4}_{11})^{m^{\prime}_{11}}(v^5_{11}v^5_{12})^{k_{22}+m^{\prime}_{11}}}{m^{\prime}_{11}!(m_{22}-m^{\prime}_{11})!}\\
&=\sum\frac{(p_3)_{m_1+m_{11}+m_{22}}(p_2+h_2)_{m_1+m_{12}}(-h_4+m_1)_{m_{11}}(-h_4)_{m_1}(\bar{p}_4+\bar{h}_4)_{m_1+m_{12}+m_{22}}}{(\bar{p}_4+\bar{h}_3)_{2m_1+m_{11}+m_{12}+m_{22}}(\bar{p}_4+\bar{h}_3+1-d/2)_{m_1}}\\
&\phantom{=}\qquad\times\binom{m_{22}-m^{\prime}_{11}}{k_{22}}\binom{k_{22}}{r_{11}}\binom{k_{22}}{r_{12}}\frac{(-1)^{k_{22}+r_{11}+r_{12}}(v^5_{11}v^5_{12})^{m^{\prime}_{11}}}{(m_{22}-m^{\prime}_{11})!}\\
&\phantom{=}\qquad\times\frac{(u^{\prime 4}_1v^5_{11}v^5_{12})^{m_1}}{m_1!}\frac{(1-v^5_{11})^{m_{11}+r_{11}}}{m_{11}!}\frac{(1-v^5_{12})^{m_{12}+r_{12}}}{m_{12}!}\frac{(1-v^{\prime 4}_{11})^{m^{\prime}_{11}}}{m^{\prime}_{11}!}.
}
We change $k_{22}$ again by $k_{22}\rightarrow k_{22}+r_{11}$ such that $G_5$ becomes
\eqna{
G_5&=\sum\frac{(p_3)_{m_1+m_{11}+m_{22}}(p_2+h_2)_{m_1+m_{12}}(-h_4+m_1)_{m_{11}}(-h_4)_{m_1}(\bar{p}_4+\bar{h}_4)_{m_1+m_{12}+m_{22}}}{(\bar{p}_4+\bar{h}_3)_{2m_1+m_{11}+m_{12}+m_{22}}(\bar{p}_4+\bar{h}_3+1-d/2)_{m_1}}\\
&\phantom{=}\qquad\times\binom{m_{22}-m^{\prime}_{11}-r_{11}}{k_{22}}\binom{k_{22}+r_{11}}{r_{12}}\frac{(-1)^{k_{22}+r_{12}}(v^5_{11}v^5_{12})^{m^{\prime}_{11}}}{(m_{22}-m^{\prime}_{11}-r_{11})!r_{11}!}\\
&\phantom{=}\qquad\times\frac{(u^{\prime 4}_{1}v^5_{11}v^5_{12})^{m_1}}{m_1!}\frac{(1-v^5_{11})^{m_{11}+r_{11}}}{m_{11}!}\frac{(1-v^5_{12})^{m_{12}+r_{12}}}{m_{12}!}\frac{(1-v^{\prime 4}_{11})^{m^{\prime}_{11}}}{m^{\prime}_{11}!}.
}
With the help of the following identity,
\eqn{(1-v)^{a}=\sum_{i}\binom{a}{i}(-1)^iv^i,}[EqIdBinom]
we can evaluate the summation over $r_{12}$ and we find
\eqna{
G_5&=\sum\frac{(p_3)_{m_1+m_{11}+m_{22}}(p_2+h_2)_{m_1+m_{12}}(-h_4+m_1)_{m_{11}}(-h_4)_{m_1}(\bar{p}_4+\bar{h}_4)_{m_1+m_{12}+m_{22}}}{(\bar{p}_4+\bar{h}_3)_{2m_1+m_{11}+m_{12}+m_{22}}(\bar{p}_4+\bar{h}_3+1-d/2)_{m_1}}\\
&\phantom{=}\qquad\times\binom{m_{22}-m^{\prime}_{11}-r_{11}}{k_{22}}\frac{(-1)^{k_{22}}(v^5_{12})^{k_{22}+r_{11}}(v^5_{11}v^5_{12})^{m^{\prime}_{11}}}{(m_{22}-m^{\prime}_{11}-r_{11})!r_{11}!}\\
&\phantom{=}\qquad\times\frac{(u^{\prime 4}_1v^5_{11}v^5_{12})^{m_1}}{m_1!}\frac{(1-v^5_{11})^{m_{11}+r_{11}}}{m_{11}!}\frac{(1-v^5_{12})^{m_{12}}}{m_{12}!}\frac{(1-v^{\prime 4}_{11})^{m^{\prime}_{11}}}{m^{\prime}_{11}!}.
}
In a similar way, we evaluate the sum over $k_{22}$ and the result is
\eqna{
G_5&=\sum\frac{(p_3)_{m_1+m_{11}+m_{22}}(p_2+h_2)_{m_1+m_{12}}(-h_4+m_1)_{m_{11}}(-h_4)_{m_1}(\bar{p}_4+\bar{h}_4)_{m_1+m_{12}+m_{22}}}{(\bar{p}_4+\bar{h}_3)_{2m_1+m_{11}+m_{12}+m_{22}}(\bar{p}_4+\bar{h}_3+1-d/2)_{m_1}}\\
&\phantom{=}\qquad\times\frac{(v^5_{12})^{r_{11}}(v^5_{11}v^5_{12})^{m^{\prime}_{11}}}{(m_{22}-m^{\prime}_{11}-r_{11})!r_{11}!}\frac{(u^{\prime 4}_1v^5_{11}v^5_{12})^{m_1}}{m_1!}\frac{(1-v^5_{11})^{m_{11}+r_{11}}}{m_{11}!}\frac{(1-v^5_{12})^{m_{12}+m_{22}-m^{\prime}_{11}-r_{11}}}{m_{12}!}\frac{(1-v^{\prime 4}_{11})^{m^{\prime}_{11}}}{m^{\prime}_{11}!}.
}
Now, we change $m_{22}$ by $m_{22}\rightarrow m_{22}+m^{\prime}_{11}+r_{11}$ and define new variables
\eqn{n_{11}=m_{11}+r_{11},\qquad n_{12}=m_{12}+m_{22},}
leading to
\eqna{
G_5&=\sum\frac{(p_3)_{m_1+m^{\prime}_{11}+n_{11}+n_{12}-m_{12}}(p_2+h_2)_{m_1+m_{12}}(-h_4+m_1)_{m_{11}}(-h_4)_{m_1}(\bar{p}_4+\bar{h}_4)_{m_1+m^{\prime}_{11}+n_{11}+n_{12}-m_{11}}}{(\bar{p}_4+\bar{h}_3)_{2m_1+m^{\prime}_{11}+n_{11}+n_{12}}(\bar{p}_4+\bar{h}_3+1-d/2)_{m_1}}\\
&\phantom{=}\qquad\times\frac{(v^5_{12})^{n_{11}-m_{11}}(v^5_{11}v^5_{12})^{m^{\prime}_{11}}}{(n_{12}-m_{12})!(n_{11}-m_{11})!}\frac{(u^{\prime 4}_1v^5_{11}v^5_{12})^{m_1}}{m_1!}\frac{(1-v^5_{11})^{n_{11}}}{m_{11}!}\frac{(1-v^5_{12})^{n_{12}}}{m_{12}!}\frac{(1-v^{\prime 4}_{11})^{m^{\prime}_{11}}}{m^{\prime}_{11}!}.
}
We then proceed to sum over $m_{12}$.  The relevant terms are
\eqn{
\begin{gathered}
(p_3)_{m_1+m^{\prime}_{11}+n_{11}+n_{12}-m_{12}}=\frac{(-1)^{m_{12}}(p_3)_{m_1+m^{\prime}_{11}+n_{11}+n_{12}}}{(1-p_3-m_1-m^{\prime}_{11}-n_{11}-n_{12})_{m_{12}}},\\
(p_2+h_2)_{m_1+m_{12}}=(p_2+h_2)_{m_1}(p_2+h_2+m_1)_{m_{12}},\\
\frac{1}{(n_{12}-m_{12})!}=\frac{(-1)^{m_{12}}(-n_{12})_{m_{12}}}{n_{12}!},
\end{gathered}
}
hence the sum over $m_{12}$ leads to the hypergeometric function 
\eqn{{}_2F_1\left[\begin{array}{c}-n_{12},p_2+h_2+m_1\\1-p_3-m_1-m^{\prime}_{11}-n_{11}-n_{12}\end{array};1\right].}
With the help of the identity ${}_2F_1(-m,b;c;1)=\frac{(c-b)_m}{(c)_m}$, we find
\eqna{
G_5&=\sum\frac{(p_3)_{m_1+m^{\prime}_{11}+n_{11}}(p_2+h_2)_{m_1}(-h_4+m_1)_{m_{11}}(-h_4)_{m_1}(\bar{p}_4+\bar{h}_4)_{m_1+m^{\prime}_{11}+n_{11}+n_{12}-m_{11}}}{(\bar{p}_4+\bar{h}_3)_{2m_1+m^{\prime}_{11}+n_{11}}(\bar{p}_4+\bar{h}_3+1-d/2)_{m_1}}\\
&\phantom{=}\qquad\times\frac{(v^5_{12})^{n_{11}-m_{11}}(v^5_{11}v^5_{12})^{m^{\prime}_{11}}}{(n_{11}-m_{11})!}\frac{(u^{\prime 4}_1v^5_{11}v^5_{12})^{m_1}}{m_1!}\frac{(1-v^5_{11})^{n_{11}}}{m_{11}!}\frac{(1-v^5_{12})^{n_{12}}}{n_{12}!}\frac{(1-v^{\prime 4}_{11})^{m^{\prime}_{11}}}{m^{\prime}_{11}!}.
}
Using \eqref{EqIdBinom}, we can now evaluate the sum over $n_{12}$ to write
\eqna{
G_5&=\sum\frac{(p_3)_{m_1+m^{\prime}_{11}+n_{11}}(p_2+h_2)_{m_1}(-h_4+m_1)_{m_{11}}(-h_4)_{m_1}(\bar{p}_4+\bar{h}_4)_{m_1+m^{\prime}_{11}+n_{11}-m_{11}}}{(\bar{p}_4+\bar{h}_3)_{2m_1+m^{\prime}_{11}+n_{11}}(\bar{p}_4+\bar{h}_3+1-d/2)_{m_1}}\\
&\phantom{=}\qquad\times\frac{(v^5_{12})^{-\bar{p}_4-\bar{h}_4}(v^5_{11})^{m^{\prime}_{11}}}{(n_{11}-m_{11})!}\frac{(u^{\prime 4}_1v^5_{11})^{m_1}}{m_1!}\frac{(1-v^5_{11})^{n_{11}}}{m_{11}!}\frac{(1-v^{\prime 4}_{11})^{m^{\prime}_{11}}}{m^{\prime}_{11}!}.
}

Finally, we perform the sums over $m_{11}$ and $n_{11}$, which are similar to the sums over $m_{12}$ and $n_{12}$, respectively.  We then find
\eqn{G_5=(v^5_{11})^{-p_3}(v^5_{12})^{-\bar{p}_4-\bar{h}_4}\sum\frac{(p_3)_{m_1+m^{\prime}_{11}}(p_2+h_2)_{m_1}(-h_4)_{m_1}(\bar{p}_4+\bar{h}_4)_{m_1+m^{\prime}_{11}}}{(\bar{p}_4+\bar{h}_3)_{2m_1+m^{\prime}_{11}}(\bar{p}_4+\bar{h}_3+1-d/2)_{m_1}}\frac{(u^{\prime 4}_1)^{m_1}}{m_1!}\frac{(1-v^{\prime 4}_{11})^{m^{\prime}_{11}}}{m^{\prime}_{11}!},}
and thus, under the limit $\mathcal{O}_{i_4}(\eta_4)\rightarrow\mathds{1}$, we prove that
\eqn{G_5^{(d,\boldsymbol{h};\boldsymbol{p})}(\boldsymbol{u}^5,\textbf{v}^5)\rightarrow(v^5_{11})^{-p_3}(v^5_{12})^{-\bar{p}_4-\bar{h}_4}G_4^{(d,\boldsymbol{h}^{\prime};\boldsymbol{p}^{\prime})}(\boldsymbol{u}^{\prime 4},\textbf{v}^{\prime 4}).
}[EqG5]
Combining \eqref{EqPrefactor} and \eqref{EqG5}, we conclude that
\eqn{I_{5(\Delta_{k_1},\Delta_{k_2})}^{(\Delta_{i_2},\ldots,\Delta_{i_5},\Delta_{i_1})}=L_5(u^5_1)^{\Delta_{k_1}/2}(u^5_2)^{\Delta_{k_2}/2}G_5\to I_{4(\Delta_{k_1})}^{(\Delta_{i_2},\Delta_{i_3},\Delta_{i_5},\Delta_{i_1})}=L^{\prime}_4(u^{\prime 4}_1)^{\Delta_{k_1}/2}G^{\prime}_4,}
in the limit of unit operator $\mathcal{O}_{i_4}(\eta_4)\rightarrow\mathds{1}$, as claimed.


\bibliography{MPtFcts}

\end{document}